\documentclass[prd,twocolumn,amsmath,amssymb,superscriptaddress
]{revtex4-2}

\usepackage{graphicx}
\usepackage{subcaption}
\usepackage{bm}
\usepackage{amsmath,amssymb}
\usepackage{hyperref}
\hypersetup{colorlinks=true, linkcolor=blue, citecolor=blue, urlcolor=blue}
\usepackage{mathrsfs}
\usepackage{physics}
\usepackage{mathpazo}

\renewcommand{\phi}{\varphi}
\renewcommand{\theta}{\vartheta}

\begin{document}

\captionsetup{justification=justified,singlelinecheck=false}

\makeatletter
\long\def\@makecaption#1#2{%
  \vskip\abovecaptionskip
  \begingroup
  \leftskip=0pt \rightskip=0pt \parfillskip=0pt plus 1fil\relax
  \noindent #1: #2\par
  \endgroup
  \vskip\belowcaptionskip}
\makeatother

\title{Double null coordinates for Kerr-Vaidya spacetime}
\author{Xiao-Kan Guo}
 \email{kankuohsiao@whu.edu.cn}
\affiliation{School of Mathematics \& Physics, Yancheng Institute of Technology, Yancheng 224051, Jiangsu, China}
\author{Xiankai Pang}
 \email{xkpang@cwnu.edu.cn}
 \affiliation{School of Physics and Astronomy, China West Normal University, Nanchong 637009, Sichuan, China}
\date{\today}

\begin{abstract}
We construct a pair of null coordinates for the Kerr-Vaidya spacetime, a nonstationary rotating black-hole geometry whose mass varies with the retarded or advanced time. The metric is naturally written in Eddington-Finkelstein-type coordinates, but the corresponding time coordinate fails to be null when the spin is nonzero, so each null coordinate must acquire a correction of the Carter type that depends on the polar angle. We show that the outgoing and ingoing null congruences require two independent Carter-type functions. The canonical null one-forms obtained from the eikonal equation are not closed when the mass varies, and their closure conditions are different equations. For constant mass both closure equations reduce to the equation found by Arga\~naraz and Moreschi. We derive the two master equations together with the two integrating-factor equations, solve both sectors numerically, and obtain the Kerr-Vaidya metric in double-null form. We also show that the unextended null coordinates are singular at the horizon, where the advanced coordinate diverges logarithmically and the metric determinant vanishes. We construct the corresponding Kruskal-type extension with a time-dependent surface gravity and show that it is regular at the horizon.
\end{abstract}

%\keywords{Kerr--Vaidya metric, double null coordinates, Carter constant, numerical relativity}

\maketitle

\section{Introduction}\label{sec:intro}
General relativity is invariant under coordinate transformations, but a suitable choice of coordinates on a given patch of the spacetime manifold is often essential for explicit calculations. For instance, for the Schwarzschild spacetime, one can transform the spherical coordinates to the coordinate system in which a radially infalling null particle has a constant coordinate reading, which is known as the ingoing Eddington--Finkelstein coordinates \cite{Carroll}. Likewise, we can transform to the outgoing Eddington--Finkelstein coordinates adapted to the outgoing null geodesics. The Eddington--Finkelstein coordinates extend to the interior of a Schwarzschild black hole and resolve the singularity at the horizon found in the spherical coordinates. The maximally extended coordinate system for the whole Schwarzschild spacetime is given by the Kruskal--Szekeres coordinates \cite{Carroll}. A notable feature of the Kruskal--Szekeres coordinates is that the two resulting coordinates are both null.

Extending the double null coordinates to Kerr spacetime is not an easy task. Since the null geodesics in the Kerr spacetime are `twisted' by the rotation, the coordinate transformations found in the Schwarzschild case cannot be applied now. There have been various attempts to define null coordinates for the Kerr spacetime \cite{Carter66,PI98,FL03,Hay04,BV06,Arganaraz2021}. The earlier attempts either work for only some patches of the Kerr spacetime or contain divergent behaviours. The recent works by Arga\~naraz and Moreschi (AM) \cite{Arganaraz2021} find a way of constructing smooth double-null coordinates for the Kerr spacetime, although involving numerical solutions. The construction requires a particular class of radial null geodesics, that is, a hypersurface orthogonal null geodesic congruence. The corresponding constant of motion $K$ is determined numerically, and it is not a (Carter) constant but a function of $(r,\theta)$, so that the null coordinate functions can be defined in terms of $K$. These double-null coordinates have been successfully used for studying the evolution of massless scalar fields in the Kerr spacetime \cite{Arganaraz2022} and constructing the global conformal diagram for the Kerr spacetime \cite{Arganaraz2023}.

Realistic black holes evolve in time through accretion or Hawking evaporation. The Kerr-Vaidya metrics~\cite{Murenbeeld1970,VP73,Carmeli1977,Carmeli82} are the simplest non-stationary extensions of the Kerr spacetime. The Kerr-Vaidya metrics are obtained from the Vaidya metric through the Newman-Janis transformation, and carry a time-dependent mass $M(u)$ together with the spin parameter $a$. Their properties have recently been clarified in depth, cf.\ \cite{Dahal2020,Dahal2021,Dahal2025,Tan2023,DV24,GKB26}. In particular, it is shown in \cite{Dahal2020} that for $a\neq0$ the energy-momentum tensor violates the null energy condition, that the apparent horizon of the \emph{outgoing} Kerr--Vaidya metric coincides with that of the Kerr black hole whereas that of the \emph{ingoing} metric does not, and that a spherically symmetric accreting black hole produces a firewall, a region of divergent energy density, pressure and flux as perceived by an infalling observer, and that this feature persists in the outgoing Kerr--Vaidya metric. This divergent behaviour at the apparent horizon is reminiscent of the coordinate artifacts encountered in the Kerr case, and also calls for a double null coordinate system in which the causal structure is manifest.

For the spherically symmetric Vaidya metric, the search for a double null coordinate system has been a long endeavour \cite{LSM65, Waugh1986,GS04,ACS06,Saa07}, but an analytic coordinate transformation cannot be found. Waugh and Lake~\cite{Waugh1986} already showed that, in a supposed double null coordinate system, the Einstein equations reduce to a first-order nonlinear partial differential equation, whose integration introduces a free function of the null variable. A special choice of this function is required to regularize the metric across horizons. In view of the Kerr case mentioned above, we expect that a generalized study for the Kerr-Vaidya spacetime would involve a function such as the Carter-type function.

The present work aims to extend the AM construction to the Kerr-Vaidya spacetime. Two structural facts determine the construction. First, in the Eddington--Finkelstein coordinates the inverse metric satisfies $g^{uu}\neq0$ for $a\neq0$, so the Eddington--Finkelstein retarded time is \emph{not} null and the null coordinate must acquire a $\theta$-dependent Carter correction. Second, and this is the new element of the present work, when $M'\neq0$ the outgoing and ingoing canonical null one-forms satisfy \emph{different} closure conditions. Consequently a single Carter-type function cannot serve both congruences, and one needs two independent functions $K(u,r,\theta)$ and $\tilde K(u,r,\theta)$, together with two integrating factors, one for each null one-form. For constant mass both closure equations reduce to the Arga\~naraz--Moreschi equation \cite{Arganaraz2021}.

The paper is organized as follows. Section~\ref{sec:vaidya} reviews the spherical Vaidya sector \cite{Waugh1986}, which fixes the structure of the integrating factor. Section~\ref{sec:metric} presents the Kerr-Vaidya metric in Eddington-Finkelstein coordinates and its physical properties. Section~\ref{sec:eikonal} derives the eikonal equation and the two canonical null one-forms. Section~\ref{sec:master} contains the two master equations and the integrating-factor quadratures, and the numerical solutions of both sectors are described in Section~\ref{sec:numerics}. If the two Carter-type functions and the two integrating factors are known, we obtain the explicit form of the Kerr-Vaidya metric in the double null coordinates, which is given in Sec.~\ref{sec:dn}. We conclude in Sec.~\ref{sec:conclusion}.

\section{Double null coordinates for spherical spacetime}\label{sec:vaidya}
We first recall the spherical double null construction of Ref.~\cite{Waugh1986}, so as to introduce the structure of the integrating factor.

Suppose the double null coordinates $(u,v)$ exist and have been found.
The spherically symmetric metric in double null coordinates is then
\begin{equation}
\dd s^{2}=-2f(u,v)\,\dd u\,\dd v+r^{2}(u,v)\,\dd\Omega^{2},
\label{eq:wl-metric}
\end{equation}
with $g^{uv}=g^{vu}=-1/f$, $g^{uu}=g^{vv}=0$. From the Christoffel symbols
$\Gamma^{u}_{uu}=f_{u}/f$, $\Gamma^{u}_{\theta\theta}=rr_{v}/f$, $\Gamma^{\theta}_{u\theta}=r_{u}/r$, and
their $u\leftrightarrow v$ counterparts, the non-vanishing Ricci components follow:
\begin{align}
&R_{uu}=2\Bigl(\frac{f_{u}r_{u}}{f}-r_{uu}\Bigr)\frac{1}{r},\qquad
R_{vv}=2\Bigl(\frac{f_{v}r_{v}}{f}-r_{vv}\Bigr)\frac{1}{r},\nonumber\\
&R_{\theta\theta}=2\frac{r_{u}r_{v}+r\,r_{uv}}{f}+1,\qquad
R_{\phi\phi}=\sin^{2}\theta\,R_{\theta\theta},\nonumber\\
&R_{uv}=\Bigl(\frac{f_{u}f_{v}}{f}-f_{uv}\Bigr)\frac{1}{f}-2\frac{r_{uv}}{r} .
\label{eq:wl-ricci}
\end{align}

The energy-momentum tensor of the high-frequency (eikonal) approximation to a unidirectional radial flow of unpolarized radiation is 
\begin{equation}T_{ab}=\frac{h}{8\pi}\,k_{a}k_{b}\end{equation} with $k_a$ null and $h$ the radiation density. Taking the flow along the $v$ direction,
$k^{a}\propto(0,1,0,0)$, so that $k_{a}\propto\partial_{u}$ ($k_{u}=-f$, $k_{v}=0$) and \emph{only the $uu$ component} 
$T_{uu}=h f^{2}/(8\pi)$ of $T_{ab}$ is non-vanishing. Since $k$ is null the trace of $T$ vanishes, $R=0$, and $G_{ab}=R_{ab}$. The Einstein equations $G_{ab}=8\pi T_{ab}$ then give 
\begin{equation}R_{vv}=2\Bigl(\frac{f_{v}r_{v}}{f}-r_{vv}\Bigr)\frac{1}{r}=0,\end{equation} which integrates to
\begin{equation}
f=2B(u)\,r_{v},
\label{eq:wl-f}
\end{equation}
with arbitrary $B(u)$, together with
\begin{equation}
R_{\theta\theta}=2\frac{r_{u}r_{v}+r\,r_{uv}}{f}+1=0 .
\label{eq:wl-theta}
\end{equation}
(With $R_{vv}=R_{\theta\theta}=R_{uv}=0$ and $R_{uu}=hf^{2}$, the remaining components of the Einstein equations are satisfied identically, since $R_{uv}$ vanishes as a consequence of \eqref{eq:wl-theta} and \eqref{eq:wl-f}.) The mass can be obtained as the Misner-Sharp mass
\begin{equation}
m(u,v)=\frac{r}{2}+\frac{r\,r_{u}r_{v}}{f},
\label{eq:wl-mass}
\end{equation}
and the radiation density appears in $R_{uu}=h f^{2}$. Eliminating $f$ through \eqref{eq:wl-f}, the mass function
\eqref{eq:wl-mass} becomes $m=r/2+r\,r_u/(2B)$, and the identity
$\partial_v m=\frac{r_v}{2}R_{\theta\theta}-\frac{r^2 r_u}{2f}R_{vv}$
shows that \eqref{eq:wl-f} together with \eqref{eq:wl-theta}
force $\partial_v m=0$, that is, $m=m(u)$. Solving for $r_u$ then
leaves the single first-order nonlinear partial differential equation for $r(u,v)$,
\begin{equation}
r_{u}(u,v)=B(u)\Bigl(\frac{2m(u)}{r}-1\Bigr),
\end{equation}
whose general solution contains an arbitrary function of $v$.
$B(u)$ must be chosen so that $f$ in \eqref{eq:wl-f} is regular across the horizons. For $m=\mathrm{const}$ the appropriate choice reproduces the Kruskal--Szekeres metric with $f\propto r^{-1}e^{-r/2m}$.

Two lessons will be carried over to the rotating case: (i) The naive transformation from radiation coordinates to double null coordinates is obstructed, and one must work ab initio in characteristic form; (ii) The regularity across horizons is controlled by a single free function of one null variable.

\section{Kerr-Vaidya metric in Eddington-Finkelstein coordinates}\label{sec:metric}

The outgoing Kerr-Vaidya metric is defined directly in retarded Eddington-Finkelstein coordinates $(u,r,\theta,\bar\varphi)$ by the Newman--Janis transform of the Vaidya metric. Equivalently, it can be obtained from the Kerr metric written in Eddington-Finkelstein form (also known as the Kerr-Eddington coordinates, cf.\ \cite{HT90})
 by replacing the constant mass with $M(u)$.

Notice that, since $\partial_{u}$ is not a Killing vector when $M'\neq0$, the Kerr-Vaidya metric
admits \emph{no} Boyer--Lindquist form. The would-be transformation $\dd t=\dd u+\dfrac{r^{2}+a^{2}}{\Delta}\dd r$
is not integrable, because
\begin{equation}
\partial_{u}\Bigl(\frac{r^{2}+a^{2}}{\Delta}\Bigr)
=\frac{2r(r^{2}+a^{2})M'(u)}{\Delta^{2}}\neq0 .
\label{eq:nonint-bl}
\end{equation}
This non-integrability forces one to work directly in the Eddington-Finkelstein coordinates (and it has the same origin as the non-closure of the angular one-form encountered in Appendix~\ref{AnnD}).

The Boyer--Lindquist form of the Kerr metric is\begin{widetext}
\begin{equation}
\dd s^{2}=
-\frac{\Delta-a^{2}\sin^{2}\theta}{\rho^{2}}\dd t^{2}
-\frac{4Mar\sin^{2}\theta}{\rho^{2}}\dd t\,\dd\phi
+\frac{\rho^{2}}{\Delta}\dd r^{2}
+\rho^{2}\dd\theta^{2}
+\frac{\Sigma\sin^{2}\theta}{\rho^{2}}\dd\phi^{2},
\label{eq:BL}
\end{equation}
with
\begin{equation}
\rho^{2}=r^{2}+a^{2}\cos^{2}\theta,\qquad
\Delta=r^{2}-2Mr+a^{2},\qquad
\Sigma=(r^{2}+a^{2})^{2}-a^{2}\Delta\sin^{2}\theta .
\label{eq:defs}
\end{equation}
For a constant mass, the Eddington-Finkelstein transformation
\begin{equation}
\dd t=\dd u+\frac{r^{2}+a^{2}}{\Delta}\dd r,\qquad
\dd\phi=\dd\bar\varphi+\frac{a}{\Delta}\dd r .
\label{eq:trans}
\end{equation}
is integrable. And, after the Eddington-Finkelstein transformation (see the Appendix \ref{AnnA} for details), with the replacement $M\to M(u)$, the outgoing Kerr-Vaidya metric is
\begin{equation}
\dd s^{2}=-\Bigl(1-\frac{2Mr}{\rho^{2}}\Bigr)\dd u^{2}-2\dd u\,\dd r
-\frac{4Mar\sin^{2}\theta}{\rho^{2}}\dd u\,\dd\bar\varphi
+2a\sin^{2}\theta\,\dd r\,\dd\bar\varphi
+\rho^{2}\dd\theta^{2}+\frac{\Sigma\sin^{2}\theta}{\rho^{2}}\dd\bar\varphi^{2},
\label{eq:kv-metric}
\end{equation}
with $M=M(u)$, $\Delta=r^{2}-2M(u)r+a^{2}$. The ingoing (accreting) case is obtained by $u\to v$,
$M(u)\to M(v)$ together with $g_{vr}\to+1$, $g_{r\bar\varphi}\to-a\sin^{2}\theta$.

The metric \eqref{eq:kv-metric} has the $(u,r,\bar\varphi)$ block
\begin{equation}
B=\begin{pmatrix}
-\frac{\Delta-a^{2}\sin^{2}\theta}{\rho^{2}} & -1 & -\frac{2Mar\sin^{2}\theta}{\rho^{2}}\\[2pt]
-1 & 0 & a\sin^{2}\theta\\[2pt]
-\frac{2Mar\sin^{2}\theta}{\rho^{2}} & a\sin^{2}\theta & \frac{\Sigma\sin^{2}\theta}{\rho^{2}}
\end{pmatrix},
\label{eq:block}
\end{equation}
whose determinant is
\begin{equation}
\det B=-\rho^{2}\sin^{2}\theta .
\label{eq:det}
\end{equation}
The inverse follows by Cramer's rule of linear algebra, and the full inverse metric is
\begin{equation}
g^{uu}=\frac{a^{2}\sin^{2}\theta}{\rho^{2}},\quad
g^{ur}=-\frac{r^{2}+a^{2}}{\rho^{2}},\quad
g^{u\bar\varphi}=\frac{a}{\rho^{2}},\quad g^{rr}=\frac{\Delta}{\rho^{2}},\quad
g^{r\bar\varphi}=-\frac{a}{\rho^{2}},\quad
g^{\theta\theta}=\frac{1}{\rho^{2}},\quad
g^{\bar\varphi\bar\varphi}=\frac{1}{\rho^{2}\sin^{2}\theta}.
\label{eq:inv-kv}
\end{equation}
The decisive entry is
\begin{equation}
g^{uu}=\frac{a^{2}\sin^{2}\theta}{\rho^{2}}>0\quad(a\neq0),
\label{eq:guu}
\end{equation}
so the Eddington-Finkelstein retarded time is \emph{not} a null coordinate for $a\neq0$, and it becomes null only for $a=0$, where the
metric reduces to the outgoing Vaidya metric. This is not special
to the time-dependent case. For the static Kerr metric as well the Eddington-Finkelstein coordinates are \emph{null-regular}, that is, finite across $\Delta=0$, but they are \emph{not} null.

\section{Eikonal equation and the two null one-forms}\label{sec:eikonal}
For an axisymmetric function $U=U(u,r,\theta)$ the null condition
$g^{\mu\nu}\partial_{\mu}U\,\partial_{\nu}U=0$ reads, using \eqref{eq:inv-kv} (the $\bar\varphi$ terms drop because $\partial_{\bar\varphi}U=0$),
\begin{equation}
\frac{a^{2}\sin^{2}\theta}{\rho^{2}}(\partial_{u}U)^{2}
-\frac{2(r^{2}+a^{2})}{\rho^{2}}\partial_{u}U\,\partial_{r}U
+\frac{\Delta}{\rho^{2}}(\partial_{r}U)^{2}
+\frac{1}{\rho^{2}}(\partial_{\theta}U)^{2}=0 .
\end{equation}
Multiplying by $\rho^{2}$ yields the eikonal equation
\begin{equation}
a^{2}\sin^{2}\theta\,(\partial_{u}U)^{2}
-2(r^{2}+a^{2})\,\partial_{u}U\,\partial_{r}U
+\Delta\,(\partial_{r}U)^{2}
+(\partial_{\theta}U)^{2}=0.
\label{eq:eikonal}
\end{equation}
Since \eqref{eq:eikonal} is homogeneous of degree two in $\partial U$, any solution is defined up to a rescaling, and the normalization $\partial_uU=1$ used below is therefore only a choice of representative, which will be relaxed by the integrating factor of Sec.~\ref{sec:master}.

Set $\partial_{u}U=1$ and regard \eqref{eq:eikonal} as an algebraic relation among $\partial_{r}U$ and $\partial_{\theta}U$. Then, by grouping the algebraic terms, we get
\begin{equation}
\bigl[(\partial_{\theta}U)^{2}+a^{2}\sin^{2}\theta\bigr]
-2(r^{2}+a^{2})\,\partial_{r}U+\Delta\,(\partial_{r}U)^{2}=0 .
\end{equation}
We define the ``Carter-type'' combination of the outgoing congruence as
\begin{equation}
K\equiv(\partial_{\theta}U)^{2}+a^{2}\sin^{2}\theta
\;\Longrightarrow\;
(\partial_{\theta}U)^{2}=K-a^{2}\sin^{2}\theta .
\label{eq:Ksep}
\end{equation}
The resulting quadratic for $\partial_{r}U$,
\begin{equation}
\Delta\,(\partial_{r}U)^{2}-2(r^{2}+a^{2})\,\partial_{r}U+K=0,
\end{equation}
has roots
\begin{equation}
\partial_{r}U=\frac{(r^{2}+a^{2})\pm\sqrt{(r^{2}+a^{2})^{2}-\Delta K}}{\Delta} .
\label{eq:roots}
\end{equation}
The \emph{outgoing} branch is the one with the minus sign. Thus the canonical null one-form of the outgoing congruence is
\begin{equation}
\omega=\dd u+A\,\dd r+B\,\dd\theta,\qquad
A=\frac{(r^{2}+a^{2})-\sqrt{(r^{2}+a^{2})^{2}-\Delta K}}{\Delta}
=\frac{K}{P+(r^{2}+a^{2})},
\quad
B=-\sigma(\theta)\sqrt{K-a^{2}\sin^{2}\theta},
\label{eq:AB}
\end{equation}
with
\begin{equation}
P\equiv\sqrt{(r^{2}+a^{2})^{2}-\Delta K},\qquad h\equiv\sqrt{K-a^{2}\sin^{2}\theta},
\label{eq:defPH}
\end{equation}
and the hemisphere sign $\sigma(\theta)=\operatorname{sign}(\cos\theta)$.

The ingoing congruence is treated in exactly the same way, but with an \emph{independent} parameter $\tilde K(u,r,\theta)$. Setting $\partial_uV=1$ and defining $\tilde K\equiv(\partial_\theta V)^2+a^2\sin^2\theta$, the plus branch of the corresponding quadratic gives the ingoing null one-form
\begin{equation}
\omega_V=\dd u+A_V\,\dd r+B_V\,\dd\theta,\qquad
A_V=\frac{(r^{2}+a^{2})+\tilde P}{\Delta}
=\frac{\tilde K}{r^{2}+a^{2}-\tilde P},
\qquad
B_V=+\sigma(\theta)\sqrt{\tilde K-a^{2}\sin^{2}\theta},
\label{eq:ABV}
\end{equation}
with
\begin{equation}
\tilde P\equiv\sqrt{(r^{2}+a^{2})^{2}-\Delta\tilde K},\qquad \tilde h\equiv\sqrt{\tilde K-a^{2}\sin^{2}\theta}.
\label{eq:defPHV}
\end{equation}
The relative sign of the $\dd\theta$ components of $\omega$ and $\omega_V$ is fixed by the requirement that, in the static limit, both one-forms be closed with the \emph{same} function, as shown in Appendix~\ref{app:ing}.

Both one-forms are null for \emph{arbitrary} $K$ and $\tilde K$. Indeed, with $\omega=(1,A,B,0)$, $g^{-1}(\omega,\omega)=0$ becomes, after multiplying by $\rho^{2}$,
\begin{equation}
a^{2}\sin^{2}\theta-2(r^{2}+a^{2})A+\Delta A^{2}+B^{2}=0 .
\label{eq:nullcheck}
\end{equation}
Since $B^{2}=K-a^{2}\sin^{2}\theta$ and $\Delta A^{2}-2(r^{2}+a^{2})A=-K$, the left-hand side of \eqref{eq:nullcheck} is $a^{2}\sin^{2}\theta-K+(K-a^{2}\sin^{2}\theta)=0$. The same computation with the ingoing branch gives $\Delta A_V^{2}-2(r^{2}+a^{2})A_V=-\tilde K$ and
\begin{equation}
a^{2}\sin^{2}\theta-2(r^{2}+a^{2})A_V+\Delta A_V^{2}+B_V^{2}=0 ,
\label{eq:nullcheckV}
\end{equation}
so that $\omega_V$ is null as well.

The radial coefficient $A$ appears to be $0/0$ at $\Delta=0$. Writing
$P\equiv\sqrt{(r^{2}+a^{2})^{2}-\Delta K}$ and rationalising, we have
\begin{equation}
A=\frac{(r^{2}+a^{2})-P}{\Delta}
=\frac{\bigl[(r^{2}+a^{2})-P\bigr]\bigl[(r^{2}+a^{2})+P\bigr]}{\Delta\bigl[(r^{2}+a^{2})+P\bigr]}
=\frac{ K}{P+(r^{2}+a^{2})}.\label{eq:Areg}
\end{equation}
At $\Delta=0$, $P=r^{2}+a^{2}$, hence $A=K/\bigl[2(r^{2}+a^{2})\bigr]$ is finite. That is, the outgoing null one-form is regular across the horizon. The ingoing coefficient, on the contrary, behaves as
\begin{equation}
A_V=\frac{(r^{2}+a^{2})+\tilde P}{\Delta}\xrightarrow{\Delta\to0}\frac{2(r^{2}+a^{2})}{\Delta},
\label{eq:AVreg}
\end{equation}
and therefore diverges at $\Delta=0$. This divergence is the familiar logarithmic behaviour of an advanced coordinate and will be discussed in Sec.~\ref{sec:dn}.

The hemisphere sign $\sigma(\theta)=\operatorname{sign}(\cos\theta)$ deserves emphasis. Since
$h=\sqrt{K-a^{2}\sin^{2}\theta}$ behaves as $h\propto|\cos\theta|$ near the equator, it has a cusp at
$\theta=\pi/2$, whereas the $\theta$-component $B=-\sigma(\theta)h$ is smooth because
$\operatorname{sign}(\cos\theta)\,|\cos\theta|=\cos\theta$. The sign is therefore \emph{essential} for the smoothness of the null one-form across the equator and is not a global convention. The same remark applies to $B_V$.

Finally, a Carter-type separation in a radiating, rotating background is not guaranteed by any symmetry, and we do not assume one. The functions $K$ and $\tilde K$ are simply the ``constant of motion'' parameters of the outgoing and ingoing null congruences. They reduce to the function of Ref.~\cite{Arganaraz2021}, which is related to Carter's constant, in the static limit, and are determined by the closure conditions of Sec.~\ref{sec:master} for a time-dependent mass. The standard Kerr--Vaidya metric does not admit a Carter constant \cite{Tan2023}, and the functions used here are not constants of motion.

\section{Closure conditions and the two master equations}\label{sec:master}
A null one-form of the type \eqref{eq:AB} is locally the gradient of a null coordinate only if it is closed. For a general one-form $\varpi=\dd u+\mathcal A\,\dd r+\mathcal B\,\dd\theta$ one has
\begin{equation}
\dd\varpi=\mathcal A_u\,\dd u\wedge\dd r+(\mathcal B_r-\mathcal A_\theta)\,\dd r\wedge\dd\theta+\mathcal B_u\,\dd u\wedge\dd\theta,
\end{equation}
and hence
\begin{equation}
\varpi\wedge\dd\varpi=\bigl[\mathcal B_r-\mathcal A_\theta-\mathcal A\,\mathcal B_u+\mathcal B\,\mathcal A_u\bigr]\dd u\wedge\dd r\wedge\dd\theta .
\label{eq:general-closure}
\end{equation}
By the Frobenius theorem, $\varpi$ admits an integrating factor, $\dd\Phi=e^{\Lambda}\varpi$, if and only if the bracket in \eqref{eq:general-closure} vanishes. This condition is manifestly invariant under multiplication of $\varpi$ by a nonvanishing function.

\subsection{The outgoing congruence}
Applying \eqref{eq:general-closure} to $(\mathcal A,\mathcal B)=(A,B)$ and using
\begin{equation}
\partial_{r}B=-\sigma\frac{\partial_{r}K}{2h},\qquad
\partial_{\theta}A=\frac{\partial_{\theta}K}{2P},\qquad
\partial_{u}B=-\sigma\frac{\partial_{u}K}{2h},\qquad
\partial_{u}A=\frac{\partial_{u}K-2rM'(u)A^{2}}{2P},
\label{eq:ingredients}
\end{equation}
one obtains, after multiplying by $2\sigma$,
\begin{equation}
\frac{\partial_{r}K}{h}+\frac{\sigma\,\partial_{\theta}K}{P}
=\frac{A\,\partial_{u}K}{h}-\frac{h\,\partial_{u}K}{P}
+\frac{2rhM'(u)A^{2}}{P}.
\label{eq:M1}
\end{equation}
The derivatives \eqref{eq:ingredients} are collected in Appendix~\ref{app:deriv}. In the static limit ($\partial_{u}K=0$, $M'=0$), Eq.~\eqref{eq:M1}
reduces to
\begin{equation}
P\,\partial_{r}K+\sigma h\,\partial_{\theta}K=0,
\label{eq:AM}
\end{equation}
which is the double null equation of Ref.~\cite{Arganaraz2021}. In the Vaidya limit ($a=0$) one has $K=h^{2}\to0$, $P=r^{2}$ and $A=B=0$, and \eqref{eq:M1} is identically satisfied.

\subsection{The ingoing congruence}
Applying \eqref{eq:general-closure} to $(\mathcal A,\mathcal B)=(A_V,B_V)$ gives, after multiplying by $2\sigma$,
\begin{equation}
\frac{\partial_{r}\tilde K}{\tilde h}+\frac{\sigma\,\partial_{\theta}\tilde K}{\tilde P}
=\frac{A_V\,\partial_{u}\tilde K}{\tilde h}+\frac{\tilde h\,\partial_{u}\tilde K}{\tilde P}
-\frac{2r\tilde hM'(u)A_V^{2}}{\tilde P}.
\label{eq:M2}
\end{equation}
The signs of the $\partial_u\tilde K$ and $M'$ terms differ from those of \eqref{eq:M1}. This is a direct consequence of the opposite signs of the $\dd\theta$ components $B=-\sigma h$ and $B_V=+\sigma\tilde h$.

In the static limit Eq.~\eqref{eq:M2} reduces to $\tilde P\partial_r\tilde K+\sigma\tilde h\partial_\theta\tilde K=0$, i.e.\ to the same equation \eqref{eq:AM}. Consequently, for $M'=0$ one has $\tilde K=K$, and the two congruences are described by a single function, as in Ref.~\cite{Arganaraz2021}.

\subsection{A single function is not sufficient when $M'\neq0$}
It is instructive to ask whether, as in the static case, a single function could satisfy both \eqref{eq:M1} and \eqref{eq:M2}. Setting $\tilde K=K$ in \eqref{eq:M2} and subtracting \eqref{eq:M1} gives
\begin{equation}
0=\frac{2K_u\Sigma}{P\Delta h}-\frac{4rhM'(\rho_1^2+P^2)}{\Delta^2P},
\qquad \Sigma\equiv\rho_1^2-\Delta a^2\sin^2\theta,\quad \rho_1\equiv r^2+a^2,
\end{equation}
i.e. (cf. Appendix \ref{app:singleK})
\begin{equation}
\partial_{u}K=\frac{2M'(u)\,r\,h^{2}\bigl(\rho_1^{2}+P^{2}\bigr)}{\Sigma\,\Delta}.
\label{eq:singleK}
\end{equation}
But \eqref{eq:M1} already determines $\partial_uK$ in terms of the spatial derivatives of $K$. Hence a single function would have to satisfy, in addition to the master equation, the algebraic constraint \eqref{eq:singleK}. This is a nontrivial condition on the spatial profile of $K$ and is satisfied identically only when $M'=0$ (in which case it reduces to $\partial_uK=0$). We conclude that for a time-dependent mass the outgoing and ingoing congruences are genuinely independent and require two functions $K$ and $\tilde K$. This is the main structural difference between the Kerr--Vaidya and the Kerr cases.

\subsection{Equations for the auxiliary functions $k$ and $\tilde k$}
With $K=a^{2}\sin^{2}\theta+k^{2}$ ($k\ge0$, $h=k$, $\partial_{r}K=2k\,\partial_{r}k$,
$\partial_{\theta}K=2a^{2}\sin\theta\cos\theta+2k\,\partial_{\theta}k$, $\partial_{u}K=2k\,\partial_{u}k$),
Eq.~\eqref{eq:M1} becomes
\begin{equation}
\partial_{r}k=\partial_{u}k\Bigl(A-\frac{k^{2}}{P}\Bigr)
+\frac{rkM'(u)A^{2}}{P}
-\frac{\sigma\bigl(a^{2}\sin\theta\cos\theta+k\,\partial_{\theta}k\bigr)}{P},
\label{eq:master-k}
\end{equation}
and, with $\xi=1/r$ ($\partial_{r}=-\xi^{2}\partial_{\xi}$), we are led to
\begin{equation}
\frac{\partial k}{\partial\xi}
=\frac{1}{\xi^{2}}\Biggl[
\frac{\sigma\bigl(a^{2}\sin\theta\cos\theta+k\,\partial_{\theta}k\bigr)}{P}
-\frac{rkM'(u)A^{2}}{P}
-\partial_{u}k\Bigl(A-\frac{k^{2}}{P}\Bigr)
\Biggr],
\label{eq:master-k-xi}
\end{equation}
where
\begin{equation}
P=\sqrt{(r^{2}+a^{2})^{2}-\Delta(a^{2}\sin^{2}\theta+k^{2})},\qquad
A=\frac{a^{2}\sin^{2}\theta+k^{2}}{P+(r^{2}+a^{2})},\qquad r=\frac{1}{\xi}.
\end{equation}
Using $P=P_{\mathrm{hat}}/\xi^{2}$ with $P_{\mathrm{hat}}=\sqrt{(1+a^{2}\xi^{2})^{2}-\xi^{4}\Delta K}$, the
right-hand side of \eqref{eq:master-k-xi} admits the manifestly finite form
\begin{equation}
\frac{\partial k}{\partial\xi}
=\frac{\sigma\bigl(a^{2}\sin\theta\cos\theta+k\,\partial_{\theta}k\bigr)}{P_{\mathrm{hat}}}
-\frac{kM'(u)K^{2}\xi^{3}}{P_{\mathrm{hat}}\bigl(P_{\mathrm{hat}}+1+a^{2}\xi^{2}\bigr)^{2}}
-\partial_{u}k\Bigl(\frac{K}{P_{\mathrm{hat}}+1+a^{2}\xi^{2}}-\frac{k^{2}}{P_{\mathrm{hat}}}\Bigr),
\label{eq:master-k-xi-finite}
\end{equation}
which is regular at $\xi=0$.

In exactly the same way, with $\tilde K=a^{2}\sin^{2}\theta+\tilde k^{2}$, Eq.~\eqref{eq:M2} gives
\begin{equation}
\partial_{r}\tilde k=\partial_{u}\tilde k\Bigl(A_V+\frac{\tilde k^{2}}{\tilde P}\Bigr)
-\frac{r\tilde kM'(u)A_V^{2}}{\tilde P}
-\frac{\sigma\bigl(a^{2}\sin\theta\cos\theta+\tilde k\,\partial_{\theta}\tilde k\bigr)}{\tilde P},
\label{eq:master-kt}
\end{equation}
and its finite $\xi$-form
\begin{equation}
\frac{\partial \tilde k}{\partial\xi}
=\frac{\sigma\bigl(a^{2}\sin\theta\cos\theta+\tilde k\,\partial_{\theta}\tilde k\bigr)}{\tilde P_{\mathrm{hat}}}
+\frac{\tilde kM'(u)\tilde K^{2}\xi^{3}}{\tilde P_{\mathrm{hat}}\bigl(1+a^{2}\xi^{2}-\tilde P_{\mathrm{hat}}\bigr)^{2}}
-\partial_{u}\tilde k\Bigl(\frac{\tilde K}{1+a^{2}\xi^{2}-\tilde P_{\mathrm{hat}}}+\frac{\tilde k^{2}}{\tilde P_{\mathrm{hat}}}\Bigr),
\label{eq:master-kt-xi-finite}
\end{equation}
with $\tilde P_{\mathrm{hat}}=\sqrt{(1+a^{2}\xi^{2})^{2}-\xi^{4}\Delta\tilde K}$. Two features distinguish \eqref{eq:master-kt-xi-finite} from \eqref{eq:master-k-xi-finite}. The signs of the $M'$ and $\tilde k^{2}$ terms are opposite, which reflects the fact that $\tilde K$ and $K$ obey different equations when $M'\neq0$. The combination $1+a^{2}\xi^{2}-\tilde P_{\mathrm{hat}}$ replaces $P_{\mathrm{hat}}+1+a^{2}\xi^{2}$ and vanishes at $\Delta=0$, as it must because $A_V$ diverges there. It vanishes also at $\xi=0$, so that \eqref{eq:master-kt-xi-finite} is not regular at spatial infinity and its integration requires an implicit treatment of the $\partial_{u}\tilde k$ term, as described in Sec.~\ref{sec:numerics}.

Both equations are supplemented by the boundary conditions
\begin{equation}
k(\xi=0,\theta)=0,\qquad
k(\xi,\theta=0)=k(\xi,\theta=\pi)=0,\qquad
k(\xi,\theta=\pi/2)=0,
\label{eq:bc}
\end{equation}
and identically for $\tilde k$. The first condition follows from asymptotic flatness ($K\to a^{2}\sin^{2}\theta$), the second from axis regularity, and the third from the smoothness of the polar component $B=-\sigma k$ across the equator, where $k\propto|\cos\theta|$ vanishes with a cusp. For $M'=0$, Eq.~\eqref{eq:master-k-xi-finite} reduces to the equation of Ref.~\cite{Arganaraz2021}, and $\tilde k=k$.

\subsection{Integrating factors}
Since $\omega$ is not closed when $M'\neq0$, we introduce an integrating factor $\lambda$ through
\begin{equation}
\dd U=\mathcal{A}\,\omega=e^{\lambda}\bigl(\dd u+A\,\dd r+B\,\dd\theta\bigr).
\label{eq:dU}
\end{equation}
Using $\dd\omega=\dd A\wedge\dd r+\dd B\wedge\dd\theta$ one has
\begin{equation}
\dd\omega=A_{u}\dd u\wedge\dd r-A_{\theta}\dd r\wedge\dd\theta+B_{u}\dd u\wedge\dd\theta+B_{r}\dd r\wedge\dd\theta,
\end{equation}
and the closure condition $\dd(e^{\lambda}\omega)=e^{\lambda}\bigl(\dd\lambda\wedge\omega+\dd\omega\bigr)=0$ gives, coefficient by coefficient,
\begin{align}
\partial_{r}B-\partial_{\theta}A
&=\partial_{\theta}\lambda\,A-\partial_{r}\lambda\,B, \label{eq:c1}\\
\partial_{u}B
&=\partial_{\theta}\lambda-\partial_{u}\lambda\,B, \label{eq:c2}\\
\partial_{u}A
&=\partial_{r}\lambda-\partial_{u}\lambda\,A . \label{eq:c3}
\end{align}
Substituting \eqref{eq:c2} and \eqref{eq:c3} into \eqref{eq:c1}, the $\partial_{u}\lambda$ terms cancel and one is left with \eqref{eq:M1}. Conversely, when \eqref{eq:M1} holds, $\lambda$ is determined by \eqref{eq:c3} up to a function of $U$, the reparametrization gauge $U\to f(U)$.

In $\xi$ the first of \eqref{eq:c3} takes the finite form
\begin{equation}
\frac{\partial\lambda}{\partial\xi}
=-\frac{k\,\partial_{u}k}{P_{\mathrm{hat}}}
+\frac{M'(u)K^{2}\xi^{3}}{P_{\mathrm{hat}}\bigl(P_{\mathrm{hat}}+1+a^{2}\xi^{2}\bigr)^{2}}
-\partial_{u}\lambda\,\frac{K}{P_{\mathrm{hat}}+1+a^{2}\xi^{2}},
\label{eq:lambda-xi}
\end{equation}
which is integrated from $\xi=0$ (gauge $\lambda=0$ at infinity) inward.

The same construction applies to the ingoing congruence, $\dd V=e^{\lambda_V}\omega_V$, with
\begin{equation}
\partial_{r}\lambda_V=\partial_{u}A_V+\partial_{u}\lambda_V\,A_V,\qquad
\partial_{\theta}\lambda_V=\partial_{u}B_V+\partial_{u}\lambda_V\,B_V ,
\label{eq:lambdaV}
\end{equation}
whose integrability is again equivalent to \eqref{eq:M2}. In $\xi$ the radial equation reads
\begin{equation}
\frac{\partial\lambda_V}{\partial\xi}
=+\frac{\tilde k\,\partial_{u}\tilde k}{\tilde P_{\mathrm{hat}}}
-\frac{M'(u)\tilde K^{2}\xi^{3}}{\tilde P_{\mathrm{hat}}\bigl(1+a^{2}\xi^{2}-\tilde P_{\mathrm{hat}}\bigr)^{2}}
-\partial_{u}\lambda_V\,\frac{\tilde K}{1+a^{2}\xi^{2}-\tilde P_{\mathrm{hat}}},
\label{eq:lambdaV-xi}
\end{equation}
again with $\lambda_V=0$ at $\xi=0$.

\begin{figure}[t]
  \centering
  \begin{subfigure}[b]{0.32\textwidth}
    \centering
    \includegraphics[width=\textwidth]{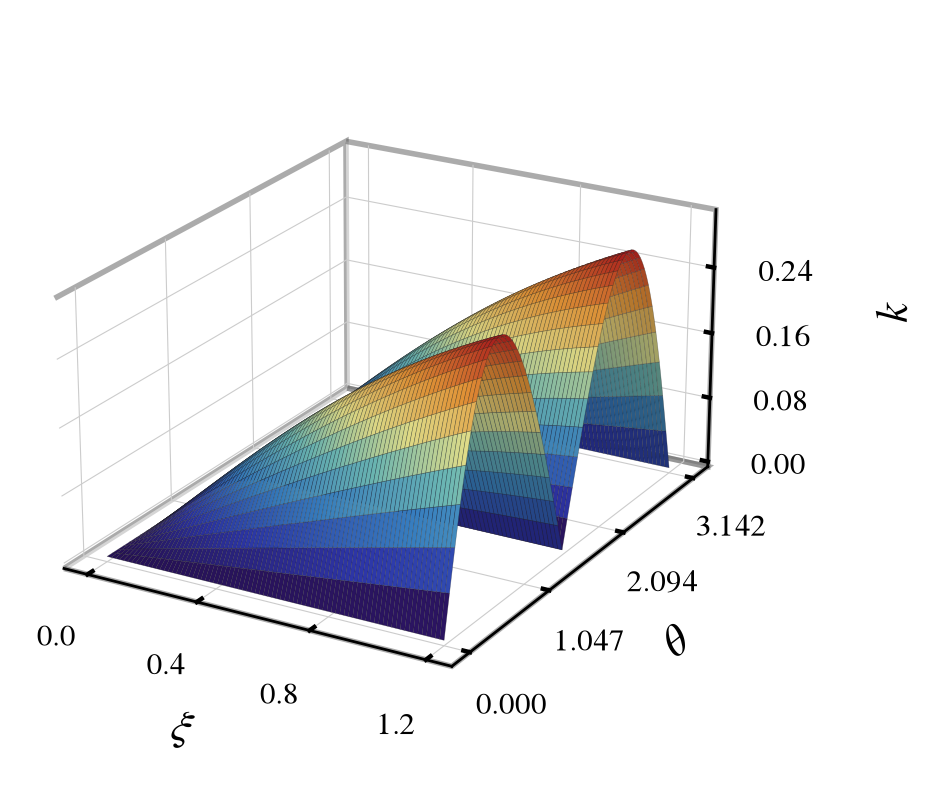}
    \caption{$k(\xi,\theta)$, $a=0.8$}
    \label{fig:k-a08}
  \end{subfigure}\hfill
  \begin{subfigure}[b]{0.32\textwidth}
    \centering
    \includegraphics[width=\textwidth]{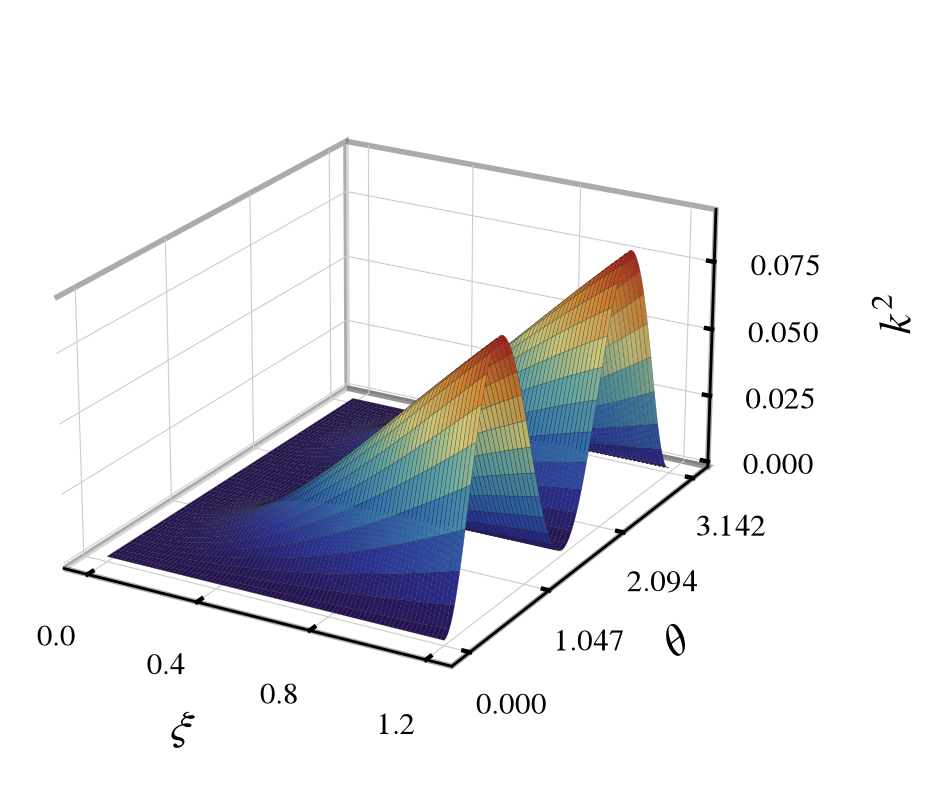}
    \caption{$k^2(\xi,\theta)$, $a=0.8$}
    \label{fig:ksq-a08}
  \end{subfigure}\hfill
  \begin{subfigure}[b]{0.32\textwidth}
    \centering
    \includegraphics[width=\textwidth]{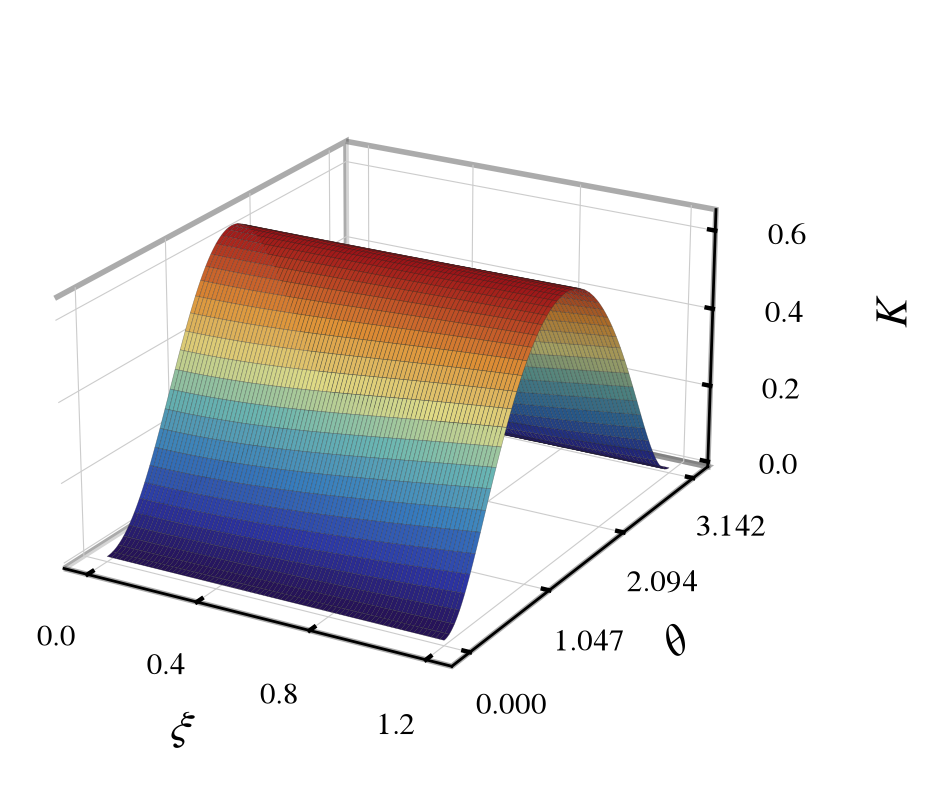}
    \caption{$K(\xi,\theta)=a^2\sin^2\theta+k^2$, $a=0.8$}
    \label{fig:K-a08}
  \end{subfigure}
  \caption{Numerical solution of the polar eikonal function $k(\xi,\theta)$,
  its square, and the function $K(\xi,\theta)=a^2\sin^2\theta+k^2$
  for the stationary Kerr geometry with spin $a=0.8$ ($M=1$), plotted against the
  compactified radius $\xi=1/r$ and the polar angle $\theta$. For $M'=0$ the
  two functions coincide, $\tilde k=k$, and these panels reproduce the
  stationary results of Ref.~\cite{Arganaraz2021}. The crease at
  $\theta=\pi/2$ is the cusp of $k\propto|\cos\theta|$ required by the
  smoothness of the null one-form across the equator.}
  \label{fig:static-a08}
\end{figure}

\begin{figure}[t]
  \centering
  \begin{subfigure}[b]{0.32\textwidth}
    \centering
    \includegraphics[width=\textwidth]{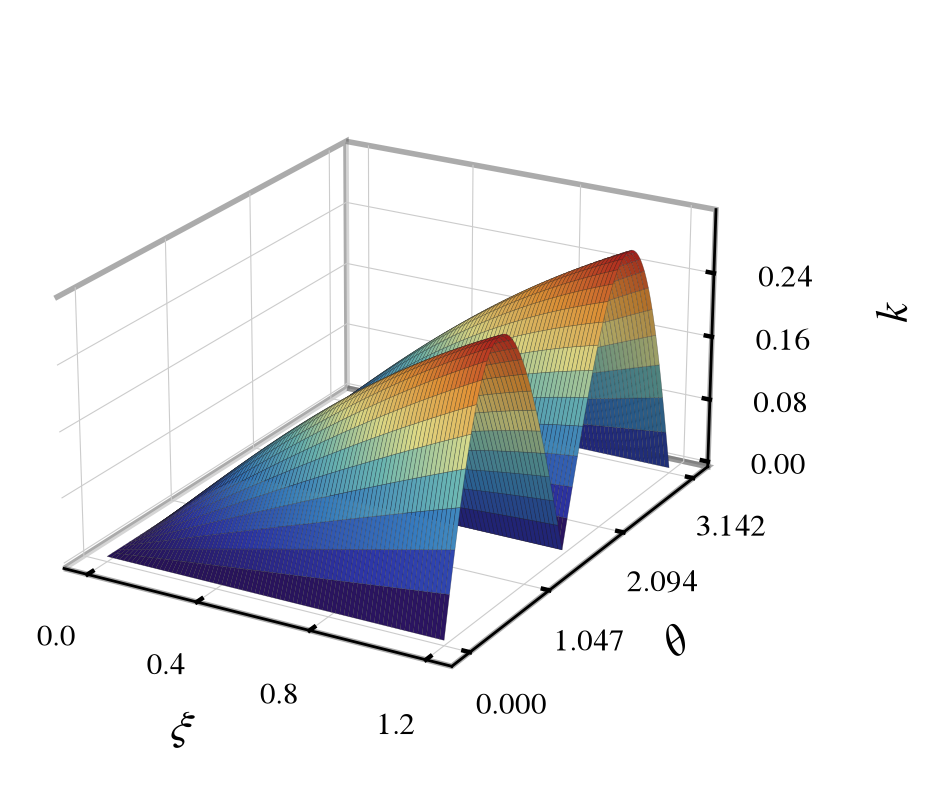}
    \caption{$k(\xi,\theta)$ at $u=50$}
    \label{fig:kv-k}
  \end{subfigure}\hfill
  \begin{subfigure}[b]{0.32\textwidth}
    \centering
    \includegraphics[width=\textwidth]{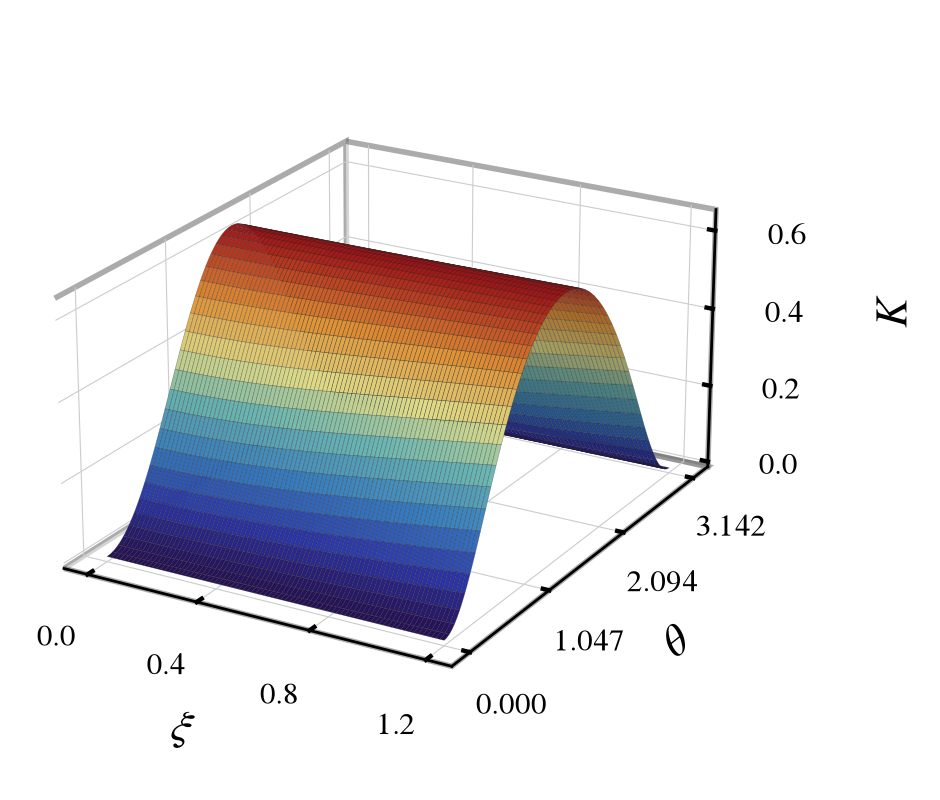}
    \caption{$K(\xi,\theta)$ at $u=50$}
    \label{fig:kv-K}
  \end{subfigure}\hfill
  \begin{subfigure}[b]{0.32\textwidth}
    \centering
    \includegraphics[width=\textwidth]{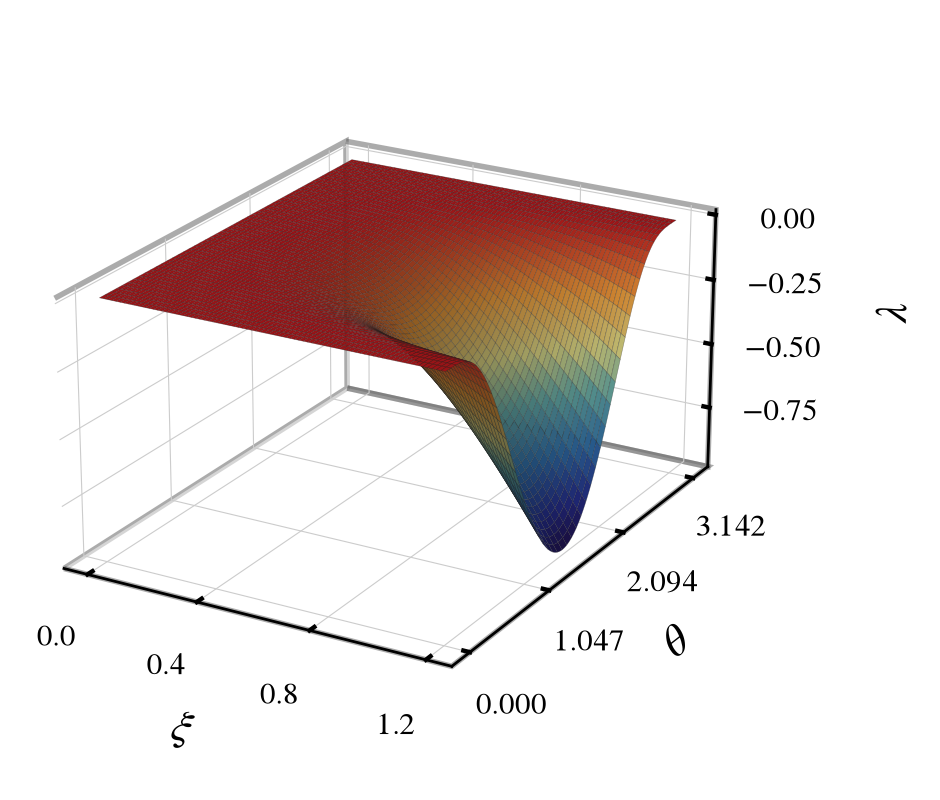}
    \caption{$\lambda(\xi,\theta)$ at $u=50$}
    \label{fig:kv-lambda}
  \end{subfigure}
  \caption{Outgoing sector of the Kerr--Vaidya spacetime, $a=0.8$, with exponential evaporation
  $M(u)=M_0 e^{-\alpha u}$ ($M_0=1$, $\alpha=0.01$), evaluated at final time
  $u=50$. Panel (a) shows the function $k$, panel (b) $K=a^2\sin^2\theta+k^2$, and
  panel (c) the integrating factor $\lambda$, in units of $10^{-4}$.}
  \label{fig:kv-evolved}
\end{figure}

\begin{figure}[t]
  \centering
  \begin{subfigure}[b]{0.32\textwidth}
    \centering
    \includegraphics[width=\textwidth]{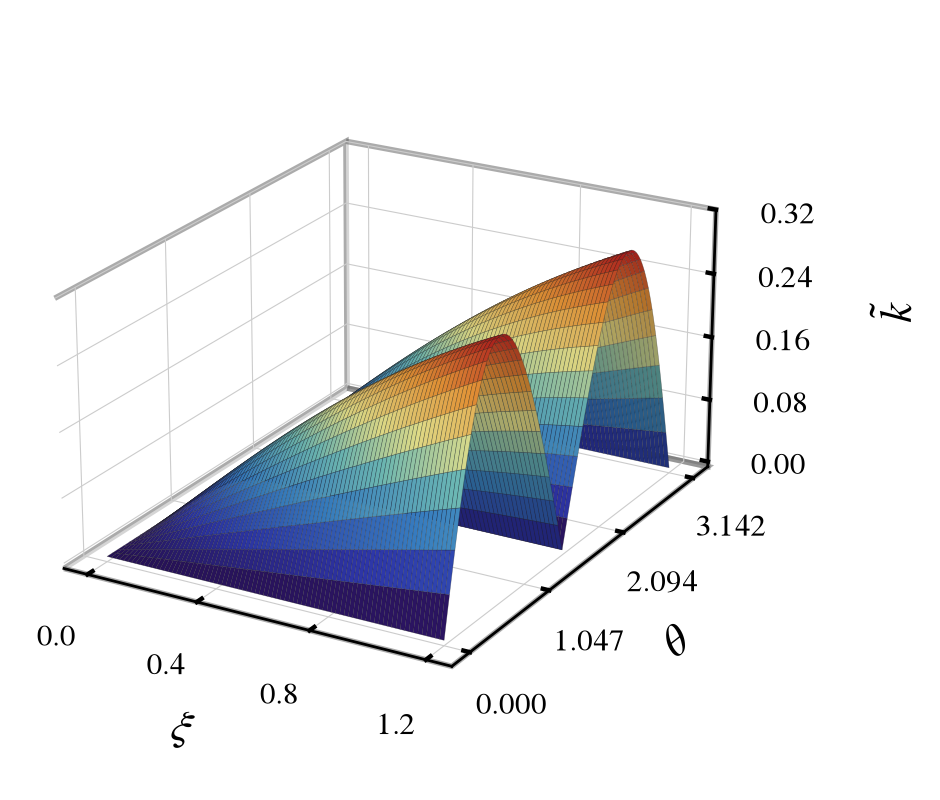}
    \caption{$\tilde k(\xi,\theta)$ at $u=50$}
    \label{fig:kvt-k}
  \end{subfigure}\hfill
  \begin{subfigure}[b]{0.32\textwidth}
    \centering
    \includegraphics[width=\textwidth]{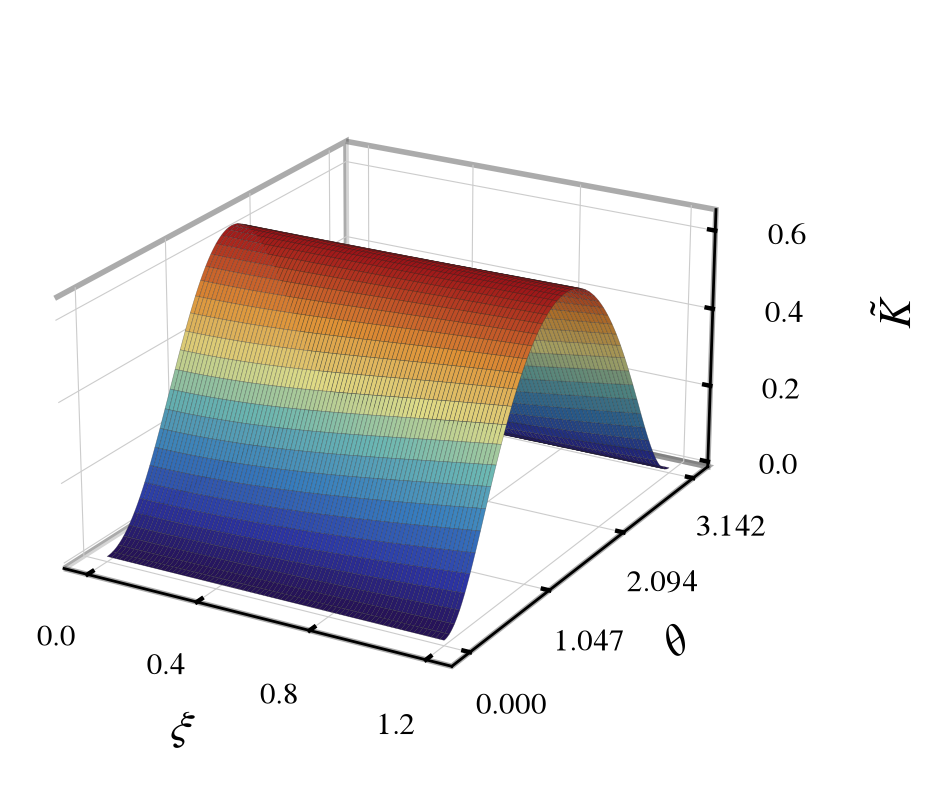}
    \caption{$\tilde K(\xi,\theta)$ at $u=50$}
    \label{fig:kvt-K}
  \end{subfigure}\hfill
  \begin{subfigure}[b]{0.32\textwidth}
    \centering
    \includegraphics[width=\textwidth]{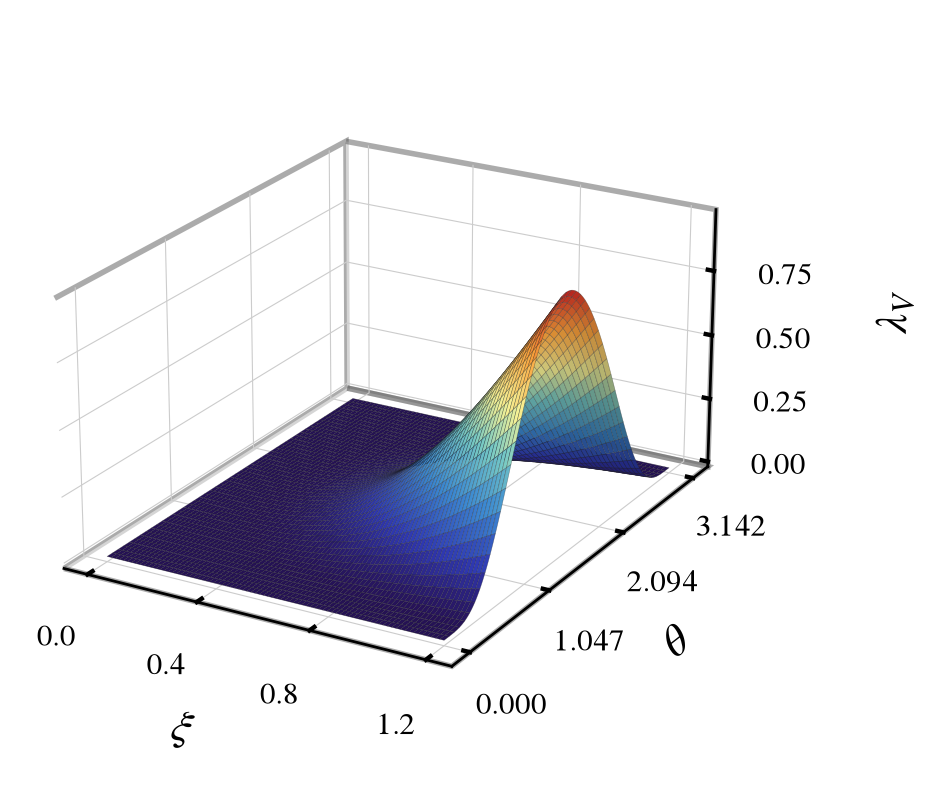}
    \caption{$\tilde k-k$ at $u=50$}
    \label{fig:kvt-diff}
  \end{subfigure}
  \caption{Ingoing sector at $u=50$ and $a=0.8$, on the domain $\xi\le0.55$,
  which corresponds to $r\ge1.82$ and lies outside the horizon at all times.
  Panel (a) shows $\tilde k$, panel (b) $\tilde K=a^2\sin^2\theta+\tilde k^2$,
  and panel (c) the difference $\tilde k-k$ with the outgoing solution of
  Fig.~\ref{fig:kv-evolved}, in units of $10^{-3}$. The two sectors almost
  coincide in the outer region and separate as the horizon is approached.}
  \label{fig:kv-ingoing}
\end{figure}

\begin{figure}[t]
  \begin{subfigure}[b]{0.42\textwidth}
    \centering
    \includegraphics[width=\textwidth]{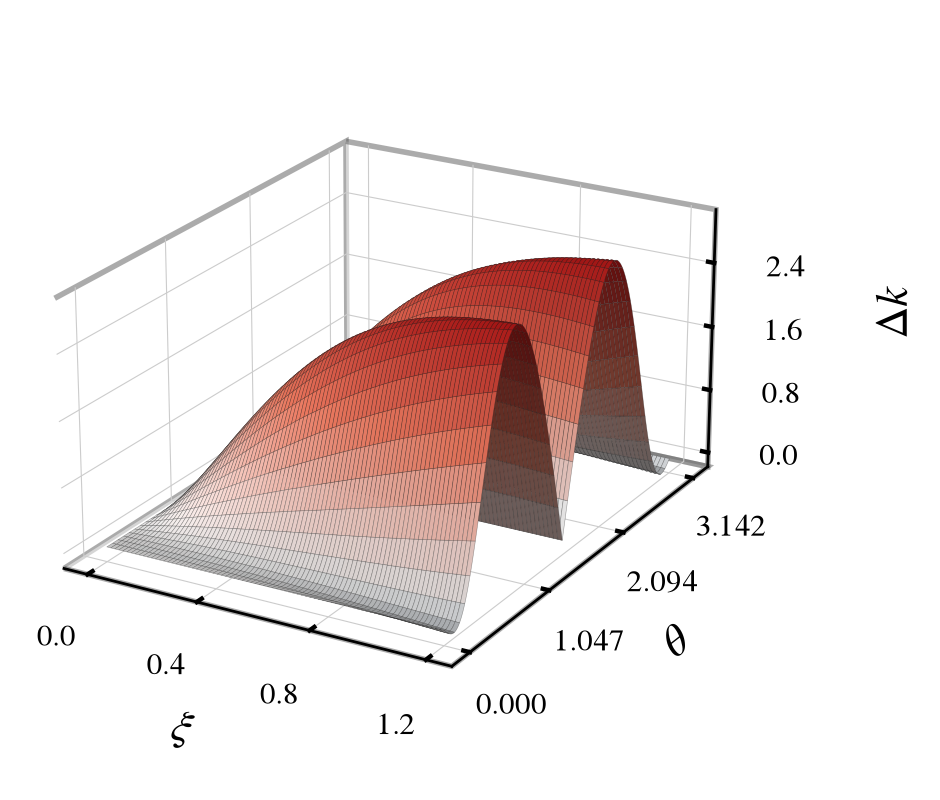}
    \caption{$\Delta k=k_{\rm KV}-k_{\rm Kerr}$}
  \end{subfigure}\hfill
  \begin{subfigure}[b]{0.42\textwidth}
    \centering
    \includegraphics[width=\textwidth]{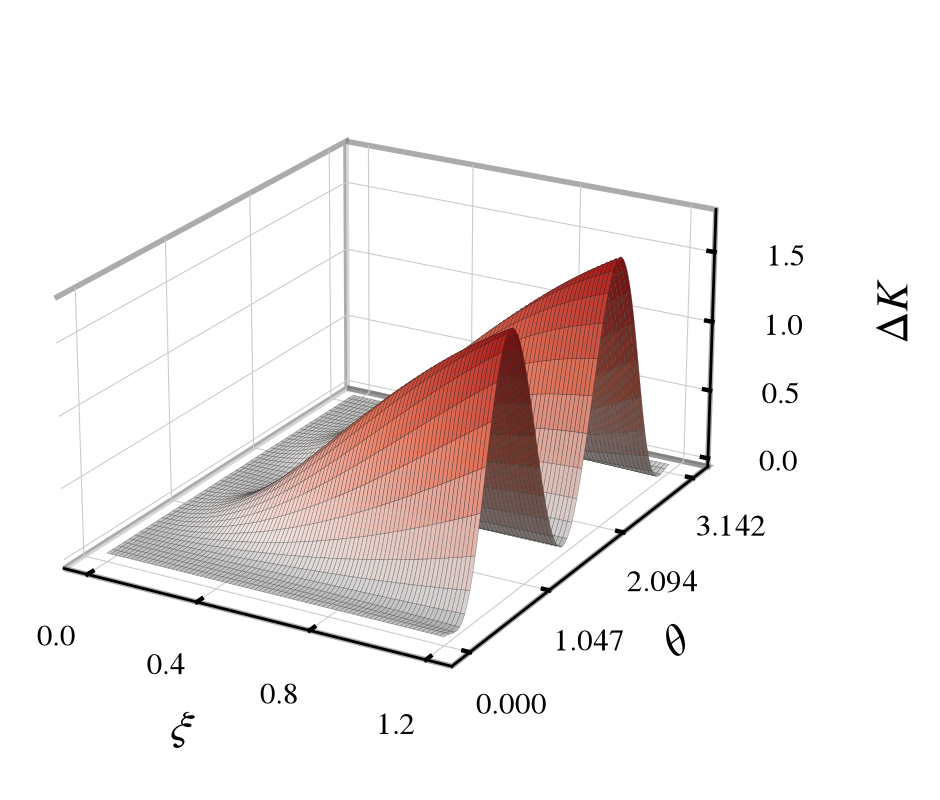}
    \caption{$\Delta K=K_{\rm KV}-K_{\rm Kerr}$}
  \end{subfigure}\hfill
  \caption{Difference between the stationary Kerr solution and the outgoing
  Kerr--Vaidya solution at $u=50$ ($a=0.8$, $M(u)=M_0e^{-\alpha u}$, $\alpha=0.01$),
  in units of $10^{-3}$.
  The reference stationary solution is evaluated with the instantaneous mass
  $M(50)$. Subtracting a solution with a different mass would instead measure the
  accumulated change of mass rather than the genuinely dynamical part.
  The diverging colour scale
  (blue $=$ negative, white $=$ zero, red $=$ positive) is symmetric about zero.}
  \label{fig:diff}
\end{figure}

\begin{figure}[t]
  \centering
  \includegraphics[width=0.85\textwidth]{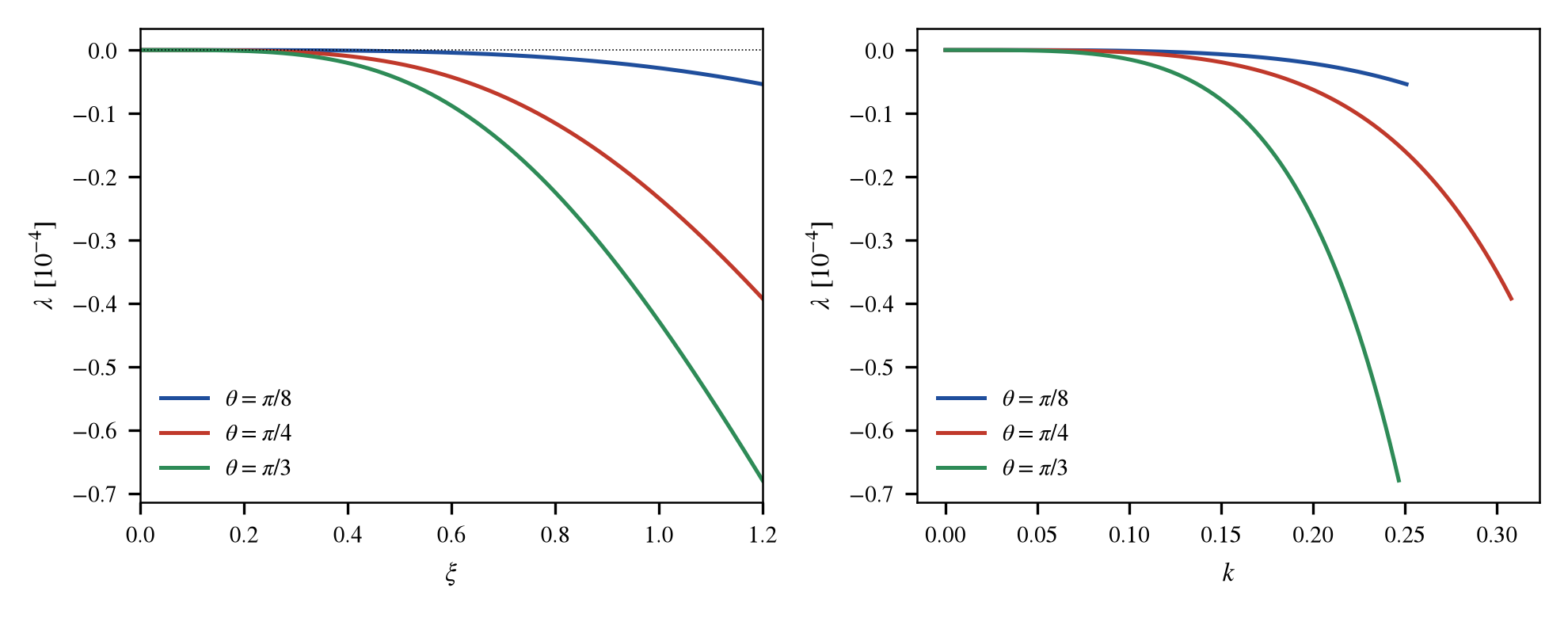}
  \caption{Integrating factor of the outgoing sector at $u=50$ and $a=0.8$:
  (a) $\lambda(\xi)$ along the radial direction for fixed polar angles
  $\theta=\pi/8,\pi/4,\pi/3$, and (b) the same data in the parametric form
  $\lambda$ versus $k$. The integrating factor vanishes at $\xi=0$
  ($r\to\infty$) and grows monotonically inward, its magnitude increasing
  towards the equator, where the Carter-type function $K$ is largest.}
  \label{fig:joint}
\end{figure}

\end{widetext}
\section{Numerical solutions}\label{sec:numerics}
The two null coordinates are determined by the two functions $k(u,\xi,\theta)$ and
$\tilde k(u,\xi,\theta)$, which satisfy the first-order master equations
\eqref{eq:master-k-xi-finite} and \eqref{eq:master-kt-xi-finite}, respectively.
We integrate the outgoing equation \eqref{eq:master-k-xi-finite} on a uniform grid in
$\xi\in[0,1.2]$ and $\theta\in[0,\pi/2]$, using the reflection symmetry about the
equator, with $N_\xi\times N_\theta=241\times 121$ points. The ingoing equation
\eqref{eq:master-kt-xi-finite} is integrated on the same $\theta$ grid and on
$\xi\in[0,0.55]$, with $221$ radial points.
Angular derivatives use second-order central differences, one-sided at the
boundaries. The radial $\xi$-integration uses a fourth-order
Runge--Kutta (RK4) step, and the time evolution employs a three-point backward
difference for $\partial_u k$ with $\Delta u=0.05$, advanced to $u_{\max}=50$.
The integrating factor $\lambda(u,\xi,\theta)$ is
reconstructed a posteriori from \eqref{eq:lambda-xi}
with the same RK4/backward-difference scheme.

The boundary conditions are $k=0$ on the symmetry axis $\theta=0,\pi$,
and at the equator
$\theta=\pi/2$, required by regularity of the polar component
$B=-\sigma k$ of the null covector \eqref{eq:AB} (with
$k$ vanishing at the equator as $|\cos\theta|$), and $k=0$ at $\xi=0$ ($r\to\infty$), so
that at spatial infinity $K\to a^2\sin^2\theta$. The same conditions are imposed on
$\tilde k$. We adopt the exponential evaporation model $M(u)=M_0e^{-\alpha u}$ with
$M_0=1$ and $\alpha=0.01$.

Figure~\ref{fig:static-a08} displays the
stationary Kerr solution for $a=0.8$ ($M=1$), which is the common initial datum
of the two sectors. Figure~\ref{fig:kv-evolved} shows the outgoing fields at $u=50$ and $a=0.8$.
At that time
$M(50)/M_0=e^{-0.5}\simeq0.607$, so that $M(50)<a$ and the spacetime is in the
naked-singularity (horizonless) regime. No horizon is present in this example, and
the integration is restricted to $r\ge0.8$.

Figure~\ref{fig:kv-ingoing} shows the corresponding ingoing fields. Because the coefficient
$\tilde K/(1+a^{2}\xi^{2}-\tilde P_{\mathrm{hat}})$ grows as $2/\xi^{2}$ at spatial infinity, the
terms proportional to $\partial_{u}\tilde k$ and $\partial_{u}\lambda_V$ are treated implicitly in
$\xi$ and the remaining terms explicitly, with the coefficients of the implicit terms evaluated
at the previous time level.
The integration is restricted to $\xi\le0.55$, which keeps $r\ge1.82$ and therefore stays outside
the horizon $r_{+}\le1.6$ at all times, since the ingoing chart is singular at $\Delta=0$. With this
scheme the static solution is preserved to $10^{-5}$ after one hundred time steps, and the
solution changes by $4\times10^{-6}$ when $\Delta u$ is halved. The two sectors agree to better than
$5\times10^{-4}$ for $\xi\lesssim0.1$, and their difference grows with $\xi$, reaching
$9\times10^{-3}$ at the inner edge of the ingoing grid. This growth is the numerical counterpart of
the divergence of $A_V$ as $\Delta\to0$. We do not display $\lambda_V$, whose equation involves the
same singular coefficient and whose value therefore depends on the gauge at $\xi=0$.

The departure from the static surfaces encodes the $M'(u)$ terms. To isolate the part of
the solution that is genuinely dynamical, in Fig.~\ref{fig:diff} we subtract the stationary
Kerr solution, evaluated with the instantaneous mass $M(50)$, from the outgoing Kerr--Vaidya solution at $u=50$. The difference fields
$\Delta k=k_{\rm KV}-k_{\rm Kerr}$ and $\Delta K=K_{\rm KV}-K_{\rm Kerr}$,
of order $10^{-3}$ and therefore a fraction of a percent of $k$ and $K$, quantify how much
the evaporation reshapes the congruence. Both $\Delta k$ and
$\Delta K$ inherit the boundary conditions $k=0$ on the axis, at the equator, and at $\xi=0$, so
the corrections are confined to the interior and peak at intermediate polar angles.

To expose the interplay between the congruence function and the integrating
factor, Fig.~\ref{fig:joint} shows $\lambda$ along the radial direction and in the
parametric form $\lambda$ versus $k$. Because $\lambda$ is driven solely by the
$M'(u)$ and $\partial_u k$ terms, it vanishes identically in
the stationary Kerr limit, so its profile is a direct map of the
evaporative dynamics. $\lambda$ vanishes at $\xi=0$, grows monotonically inward,
and is largest at the equator, where $K$ attains its maximum. This is consistent
with the source terms of Eq.~\eqref{eq:lambda-xi}. Its overall magnitude,
$|\lambda|\lesssim10^{-4}$, shows that the integrating factor is a small but
nonvanishing correction to the null one-form.

The numerical scheme is validated by the residual of the master equation
\eqref{eq:master-k-xi-finite} and of the closure
conditions \eqref{eq:c2}--\eqref{eq:c3}, which are satisfied to better than $10^{-8}$
and $10^{-6}$, respectively, over the full grid. (The null conditions
\eqref{eq:nullcheck} are algebraic identities and do not constitute an independent test.)

\section{Kerr-Vaidya metric in double-null coordinates }\label{sec:dn}
Once the two functions $K,\tilde K$ and the two integrating factors are known, the
null coordinates are determined by
\begin{equation}
\dd U=e^{\lambda}\,\omega,\qquad
\dd V=e^{\lambda_V}\,\omega_{V},
\label{eq:dU-dV}
\end{equation}
and the metric can be written in the double-null basis. In the
Eddington--Finkelstein coordinates of Sec.~\ref{sec:metric} the two
null one-forms of Secs.~\ref{sec:eikonal} are
\begin{equation}
\omega=\dd u+A\,\dd r+B\,\dd\theta,\qquad
\omega_{V}=\dd u+A_V\,\dd r+B_V\,\dd\theta,
\label{eq:omega-omegaV}
\end{equation}
where
\begin{align}
A&=K/\bigl[P+(r^{2}+a^{2})\bigr],\quad B=-\sigma k\nonumber\\
A_V&=\tilde K/\bigl[r^{2}+a^{2}-\tilde P\bigr]=\bigl[(r^{2}+a^{2})+\tilde P\bigr]/\Delta\nonumber\\
B_V&=+\sigma\tilde k
\end{align}
with
$k=\sqrt{K-a^{2}\sin^{2}\theta}$ 
and $\tilde k=\sqrt{\tilde K-a^{2}\sin^{2}\theta}$.

Solving \eqref{eq:omega-omegaV} for $\dd u$ and $\dd r$ gives
\begin{equation}
\dd r=\delta\bigl(\omega_{V}-\omega\bigr)+\varepsilon\,\dd\theta,
\qquad
\dd u=\alpha\,\omega+\beta\,\omega_{V}+\gamma\,\dd\theta,
\label{eq:app-dr}
\end{equation}
with
\begin{equation}
\begin{aligned}
\delta&=\frac{\Delta}{P+\tilde P},\qquad
\alpha=\frac{r^{2}+a^{2}+\tilde P}{P+\tilde P},\qquad
\beta=-\frac{r^{2}+a^{2}-P}{P+\tilde P},\\
\varepsilon&=\delta\,(B-B_V),\qquad
\gamma=-A\,\varepsilon-B .
\end{aligned}
\label{eq:app-coeffs}
\end{equation}
Substituting \eqref{eq:app-dr} into \eqref{eq:kv-metric} and collecting the
coefficients of $\omega^{2}$, $\omega\omega_V$, $\omega_V^{2}$, $\dd\theta^{2}$, etc.,
one obtains
\begin{widetext}
\begin{equation}
\dd s^{2}={}\mathfrak A\,\omega^{2}+2\mathfrak B\,\omega\,\omega_{V}+\mathfrak C\,\omega_{V}^{2}
+2\mathfrak D\,\omega\,\dd\theta+2\mathfrak E\,\omega_{V}\,\dd\theta+\mathfrak F\,\dd\theta^{2}+2\mathfrak G\,\omega\,\dd\bar\varphi+2\mathfrak H\,\omega_{V}\,\dd\bar\varphi
+2\mathfrak I\,\dd\theta\,\dd\bar\varphi+\mathfrak J\,\dd\bar\varphi^{2},
\label{eq:kv-dn}
\end{equation}
where, with $f\equiv1-2Mr/\rho^{2}$, $w\equiv2Mar\sin^{2}\theta/\rho^{2}$ and $j\equiv a\sin^{2}\theta$,
\begin{align}
\mathfrak A&=-f\alpha^{2}+2\alpha\delta, &
\mathfrak B&=-f\alpha\beta-\delta(\alpha-\beta), &
\mathfrak C&=-f\beta^{2}-2\beta\delta,\nonumber\\
\mathfrak D&=-f\alpha\gamma-(\alpha\varepsilon-\gamma\delta), &
\mathfrak E&=-f\beta\gamma-(\beta\varepsilon+\gamma\delta), &
\mathfrak F&=-f\gamma^{2}-2\gamma\varepsilon+\rho^{2},\nonumber\\
\mathfrak G&=-w\alpha-j\delta, &
\mathfrak H&=-w\beta+j\delta, &
\mathfrak I&=-w\gamma+j\varepsilon,\qquad
\mathfrak J=\frac{\Sigma\sin^{2}\theta}{\rho^{2}}.
\label{eq:kv-dn-coeffs}
\end{align}
\end{widetext}
Replacing $\omega\to e^{-\lambda}\dd U$ and $\omega_V\to e^{-\lambda_V}\dd V$ in
\eqref{eq:kv-dn} gives the Kerr--Vaidya metric in the coordinates
$(U,V,\theta,\bar\varphi)$. The coefficients \eqref{eq:kv-dn-coeffs} are verified
in Appendix~\ref{AnnD}, where it is also shown that $g^{UU}=g^{VV}=0$ identically,
as required for null coordinates, and that
\begin{equation}
\det g=-\frac{\rho^{4}\sin^{2}\theta\,\Delta^{2}}{(P+\tilde P)^{2}}\,
e^{-2(\lambda+\lambda_V)} .
\label{eq:det-dn}
\end{equation}
By construction $g^{UU}=g^{VV}=0$, so both $U$ and $V$ are null coordinates.
Note, however, that $g_{UU},g_{VV}\neq0$ for $a\neq0$, the $\dd U^{2},\dd V^{2}$
pieces encoding the twist of the null congruences. This is the same structure
as in \cite{Arganaraz2021}. Moreover, cross terms between the null directions and the
angular coordinates are present. In the static case they are removed by the
azimuthal transformation of Sec.~\ref{sec:metric}, which is not integrable for
$M'\neq0$.

We record three limits.
\begin{enumerate}
\item For $r\to\infty$ one has $f\to1$, $\delta\to1/2$, $\alpha\to1$, $\beta\to0$, $B,B_V\to0$, so that
$\mathfrak A\to0$, $\mathfrak B\to-1/2$, and \eqref{eq:kv-dn} reduces to
$\dd s^{2}=-\dd U\,\dd V+r^{2}\dd\Omega^{2}$, flat space in double-null form.
\item For $a\to0$ one has $K,\tilde K\to0$, $P=\tilde P\to r^{2}$, $B=B_V=0$, and the $\dd U^{2},\dd V^{2}$
and angular cross terms vanish, leaving
$\dd s^{2}=-\bigl(1-2M/r\bigr)\dd U\dd V+r^{2}\dd\Omega^{2}$, i.e.\
$-2f\dd U\dd V+r^{2}\dd\Omega^{2}$ with $f=(1-2M/r)/2$, the outgoing Vaidya
double-null form of Sec.~\ref{sec:vaidya}.
\item For $M'=0$ one has $\tilde K=K$, $\tilde P=P$, $\delta=\Delta/(2P)$,
$\varepsilon=\Delta B/P$, $\gamma=-\rho_1B/P$, and the further transformation
$\dd\varphi=\dd\bar\varphi+(a/\Delta)\dd r$, which is integrable in this case,
removes the $\dd\theta\,\dd\bar\varphi$ cross term. The metric then reduces to
Eq.~(104) of Ref.~\cite{Arganaraz2021}.
\end{enumerate}

\subsection{Behaviour at the horizon}\label{app:dn-horizon}
The chart $(U,V,\theta,\bar\varphi)$ does \emph{not} cover the horizon, and it is
important to state this precisely. In the black-hole phase $M>a$ the outer horizon is at
$r=r_+=M+\sqrt{M^{2}-a^{2}}$, where $\Delta=0$, $P=\tilde P=r^{2}+a^{2}$ and
\begin{equation}
A_V=\frac{r^{2}+a^{2}+\tilde P}{\Delta}\longrightarrow\frac{2(r^{2}+a^{2})}{\Delta} .
\label{eq:app-hor-AV}
\end{equation}
Consequently, along a curve of constant $u$ and $\theta$,
\begin{equation}
V=\int e^{\lambda_V}A_V\,\dd r\simeq\frac{1}{\kappa}\ln|\Delta|+\text{const}
\longrightarrow-\infty,
\label{eq:app-hor-V}
\end{equation}
with $\kappa=(r_+-M)/(r_+^{2}+a^{2})$ the surface gravity, i.e.\ the horizon lies at $V=-\infty$.
Correspondingly the Jacobian of the map
$(u,r)\mapsto(U,V)$ at fixed $(\theta,\bar\varphi)$ is
\begin{equation}
J=\frac{\partial(U,V)}{\partial(u,r)}=e^{\lambda+\lambda_V}\bigl(A_V-A\bigr)
=e^{\lambda+\lambda_V}\frac{P+\tilde P}{\Delta},
\label{eq:app-hor-J}
\end{equation}
which diverges at $\Delta=0$, so that by \eqref{eq:det-dn} the metric is degenerate there as
$\det g\propto\Delta^{2}\to0$. Thus, although the coefficients \eqref{eq:kv-dn-coeffs} remain finite
for $\Delta\to0$, the double-null chart degenerates at the horizon and the horizon is not a
surface of finite $V$.

This situation is like the one encountered in the Kerr case. The null coordinates
$(u,v)$ of Ref.~\cite{Arganaraz2021} are also singular at $\Delta=0$, and they obtain a
regular extension across the horizons by passing to the extended coordinates
$U_{\rm ext}=-e^{-\kappa u}$, $V_{\rm ext}=e^{\kappa v}$, which are finite at the horizon. The coordinates
$(U,V)$ constructed here are the direct Kerr--Vaidya analogues of the unextended
$(u,v)$, and the extension carries over with a time-dependent surface gravity.

The ingoing Carter function requires a separate comment, since its equation
contains the same $A_V$ and might appear to inherit the degeneracy. Near the
horizon the $M'$ term of \eqref{eq:master-kt} carries the coefficient
\begin{equation}
\frac{r\tilde kA_V^{2}}{\tilde P}\longrightarrow\frac{4r(r^{2}+a^{2})}{\Delta^{2}}\,\tilde k ,
\label{eq:kt-coeff}
\end{equation}
which does grow as $\Delta^{-2}$. This is a statement about the coefficients,
not about the solution. Retaining the dominant balance and writing
$\Delta\simeq\Delta'(r-r_+)$ with $\Delta'=2(r_+-M)$, the equation integrates to
\begin{equation}
\tilde k\propto\exp\!\Big(\frac{C}{r-r_+}\Big),\qquad
C=\frac{4r(r^{2}+a^{2})M'}{\Delta'^{2}} .
\label{eq:kt-exp}
\end{equation}
For evaporation, $M'<0$, one has $C<0$ and $\tilde k\to0$ together with all its
derivatives, so that $\tilde K\to a^{2}\sin^{2}\theta$. The level surfaces of
$V$ therefore lose their polar dependence at the horizon, which is the behaviour
expected of the advanced time. For accretion, $M'>0$, the same exponent is
positive and $\tilde k$ diverges. A direct integration of \eqref{eq:master-kt}
toward the horizon reproduces the exponent of \eqref{eq:kt-exp}. In neither case does the behaviour depend on
the chart. Since $\dd V_K/\dd r\simeq2\kappa(r^{2}+a^{2})$ is finite at
$\Delta=0$, passing to the extended coordinates rescales $\partial_r\tilde k$ by
a finite factor and cannot remove a divergence, nor create one. The obstruction
to a global chart is thus the geometric one stated above, namely $V\to-\infty$,
and not the Carter function.

For $M'=0$ the same expansion is already contained in Ref.~\cite{Arganaraz2021}.
There they isolate the $(r^2+a^2)/\Delta$ term of $\dd u$ and $\dd v$
and integrate it to logarithms of $r-r_\pm$, while the Carter function is
integrated through $\Delta=0$ without difficulty. Equations~\eqref{eq:kt-coeff}
and \eqref{eq:kt-exp} extend that statement to $M'\neq0$, where the coefficient
grows as $\Delta^{-2}$ and the solution becomes exponential, and both reduce to
the regular case as $M'\to0$.

\subsection{Kruskal-type extension}\label{app:dn-kruskal}
The degeneracy of the chart at $\Delta=0$ is removed by rescaling the two null
coordinates. We introduce
\begin{equation}
U_K=-e^{-\kappa(u)U},\qquad V_K=e^{\kappa(u)V},
\label{eq:Kruskal}
\end{equation}
with
\begin{equation}
\begin{aligned}
\kappa(u)&=\frac{r_+(u)-M(u)}{r_+(u)^2+a^2},\\
r_+(u)&=M(u)+\sqrt{M(u)^2-a^2},
\end{aligned}
\label{eq:kappa}
\end{equation}
the surface gravity of the apparent horizon in the black-hole phase $M>a$.

Two asymptotic relations fix the behaviour of the extended chart. Since
$A_V\Delta\to2(r_+^2+a^2)$ as $\Delta\to0$ and
$\Delta\simeq2\kappa(r_+^2+a^2)(r-r_+)$ near the horizon, the ingoing null
coordinate behaves as
\begin{equation}
\begin{aligned}
V&=\frac{1}{\kappa}\ln|\Delta|+V_0(u,\theta)+O(\Delta\ln|\Delta|),\\
U&=U_0(u,\theta)+O(\Delta).
\end{aligned}
\label{eq:Vlog}
\end{equation}
Consequently $V_K=e^{\kappa V_0}\Delta+O(\Delta^2)$ vanishes linearly at the
horizon, while $U_K=-e^{-\kappa U_0}$ is finite and nonvanishing there, so that the
horizon is the surface $V_K=0$ of the extended chart.

The metric is regular in this chart because the coefficients
\eqref{eq:kv-dn-coeffs} vanish at the horizon in a hierarchical way. Expanding
them in $\Delta$ along the horizon gives
\begin{equation}
\begin{aligned}
&\mathfrak A\to\frac{a^2\sin^2\theta}{\rho^2},\qquad
\mathfrak B=O(\Delta),\qquad
\mathfrak C=O(\Delta^2),\\
&\mathfrak D=O(1),\qquad
\mathfrak E=O(\Delta),
\end{aligned}
\label{eq:hor-scaling}
\end{equation}
together with $\mathfrak F,\mathfrak G,\mathfrak I,\mathfrak J=O(1)$ and
$\mathfrak H=O(\Delta)$, all with finite limits that vanish only for
$\sin\theta=0$. Substituting
\begin{equation}
\dd U=-\frac{\dd U_K}{\kappa U_K}-\frac{U\kappa'}{\kappa}\dd u,\qquad
\dd V=\frac{\dd V_K}{\kappa V_K}-\frac{V\kappa'}{\kappa}\dd u
\label{eq:dUdV-K}
\end{equation}
into \eqref{eq:kv-dn} gives, at leading order,
\begin{equation}
\begin{aligned}
g^K_{U_KU_K}&=\frac{\mathfrak A}{\kappa^2U_K^2},\qquad
g^K_{U_KV_K}=\frac{\mathfrak B}{\kappa^2U_KV_K},\\
g^K_{V_KV_K}&=\frac{\mathfrak C}{\kappa^2V_K^2},\qquad
g^K_{U_K\theta}=\frac{\mathfrak D}{\kappa U_K},\qquad
g^K_{V_K\theta}=\frac{\mathfrak E}{\kappa V_K},
\end{aligned}
\label{eq:gK}
\end{equation}
and the components involving $\bar\varphi$ follow in the same way from
$\mathfrak G$, $\mathfrak H$ and $\mathfrak I$. Every entry is finite at
$\Delta=0$, because the powers of $V_K=e^{\kappa V_0}\Delta$ in the denominators
match the orders \eqref{eq:hor-scaling} of the corresponding coefficients. The
determinant
\begin{equation}
\begin{aligned}
\det g_K&=\frac{\det g}{\kappa^4U_K^2V_K^2}\\
&=-\frac{\rho^4\sin^2\theta\,e^{-2(\lambda+\lambda_V)}}
{\kappa^4U_K^2(P+\tilde P)^2}
\left(\frac{\Delta}{V_K}\right)^2
\end{aligned}
\label{eq:det-K}
\end{equation}
is finite and nonvanishing at the horizon, since $\Delta/V_K\to e^{-\kappa V_0}$.

For $M'=0$ the surface gravity is constant and \eqref{eq:Kruskal} reduces to the
extended coordinates of Ref.~\cite{Arganaraz2021}. For $M'\neq0$ the
transformation \eqref{eq:dUdV-K} carries terms proportional to
$\kappa'=\dd\kappa/\dd u$. Their size is fixed by the behaviour of the
Eddington--Finkelstein time at fixed $U$. From
$\partial u/\partial V=-A\,\partial r/\partial V$ with
\begin{equation}
\frac{\partial r}{\partial V}=\frac{\Delta\,e^{-\lambda_V}}{\rho_1+\tilde P}=O(\Delta),
\label{eq:rV}
\end{equation}
one has $\partial u/\partial V=O(\Delta)$, so that $\dd u$ is dominated by
$\dd U$ and the terms proportional to $\kappa'$ are of relative order
$\kappa'U$ and $\kappa'V$ in the two null directions. At the past horizon $U$ is
finite and these terms are bounded. At the future horizon, where $U\to+\infty$,
they are small in the adiabatic regime $\kappa'U\ll1$, which holds for the
evaporating models of Sec.~\ref{sec:numerics} while $M>a$. The extension
\eqref{eq:Kruskal} is therefore regular at $\Delta=0$ to leading adiabatic order,
and it becomes exact when the mass is frozen.

\begin{figure}[t]
  \centering
  \includegraphics[width=\columnwidth]{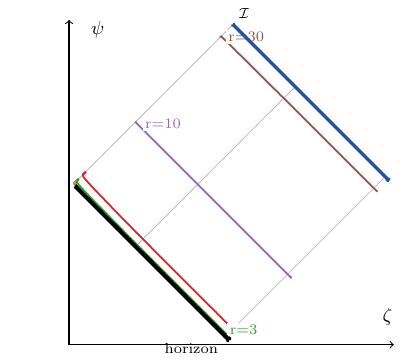}
  \caption{The exterior chart in the compactified plane \eqref{eq:CP}. Curves of
  constant $r$ are drawn in colour, curves of constant $u$ are the thin grey
  lines, the heavy black line is the horizon $V_K=0$, that is $\psi=-\zeta$, and
  the heavy blue line is the boundary $r\to\infty$, that is the null infinity
  $\mathcal{I}$. The compactification of
  Eq.~\eqref{eq:CPa} is applied to $\ln(1+|U_K|)$ and $\ln(1+V_K)$, so that the
  horizon at $V_K=0$ and infinity at $V_K\to\infty$ both fall inside the frame.
  The chart is computed from the double-null coordinates of
  Sec.~\ref{sec:numerics} with $a=0.8$, $M_0=1$, $\alpha=0.01$ over the range
  $-100\le u\le120$. The horizontal extension of a curve of constant $r$ is set
  by the variation of $U$ over the whole evolution, whereas its vertical
  extension is set only by the corresponding change of $V$, which is why the
  coloured curves stay close to the null direction. A curve of constant $r$ is
  exactly straight only for $M'=0$, where $V$ does not depend on $u$; the
  residual curvature of the coloured lines, of order two per cent of the frame,
  measures the change of $V$ at fixed $r$ during the evaporation.}
  \label{fig:chart}
\end{figure}

The chart itself is best displayed in the compactified plane
\begin{equation}
\tilde U=\arctan U_K,\qquad \tilde V=\arctan V_K,
\label{eq:CPa}
\end{equation}
\begin{equation}
\psi=\tilde V+\tilde U,\qquad \zeta=\tilde V-\tilde U,
\label{eq:CP}
\end{equation}
in which the horizon $V_K=0$ is the null line $\psi=-\zeta$ and future null
infinity $V_K\to\infty$ is the boundary $\tilde V=\pi/2$. Figure~\ref{fig:chart}
shows the exterior region obtained by integrating the null one-forms with the
numerical fields, together with the horizon. The curves of constant $r$ are
timelike and approach the horizon asymptotically, while the curves of constant
$u$ are null and run at $45^\circ$. Only the exterior is drawn, because the
chart itself stops at $\Delta=0$, where $V$ diverges.

\begin{figure}[t]
  \centering
  \includegraphics[width=0.62\columnwidth]{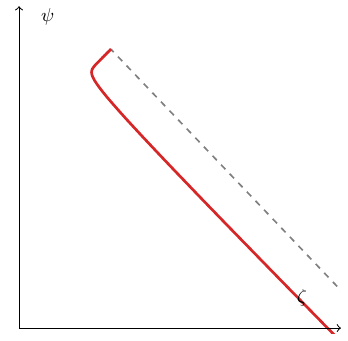}
  \caption{The bending of a curve of constant $r$, drawn in the same
  $(\psi,\zeta)$ plane as Fig.~\ref{fig:chart} and magnified. The curve is the one
  with $r=3$ and the dashed straight line is the null chord joining its end
  points. Since $\psi+\zeta=2\tilde V$, the departure of the curve from the chord
  is half the change of $\psi+\zeta$ along it, which is the quantity that would
  vanish identically for $M'=0$. The window is centred on the point of largest
  departure, where the curve is still leaving the horizon; there the departure
  reaches 0.058 in the units of the figure, so the bending is a direct, if
  small, imprint of the evaporation on the ingoing null coordinate.}
  \label{fig:bend}
\end{figure}

Figure~\ref{fig:bend} isolates the departure of these curves from the null
direction. Along a curve of constant $r$ the combination $\psi+\zeta=2\tilde V$
is twice the compactified ingoing null coordinate, so a straight curve would
mean that $V$ depends on $r$ alone, which is the case only for $M'=0$. The
figure magnifies the region where the departure is largest, which for $r=3$ is
where the curve leaves the horizon, and there the departure reaches $0.058$
before decreasing again along the curve. What is displayed is therefore the
accumulated change of the ingoing null coordinate produced by the evaporation,
that is, the imprint of the outgoing radiation on the ingoing congruence. The
effect is small in the adiabatic regime used here, of order a few per cent of
the frame of Fig.~\ref{fig:chart}, and it would grow for a faster evaporation.

The origin of the drift is the dependence of $A_V$ on the mass. In
\eqref{eq:ABV} the coefficient of $\dd r$ is
$A_V=[(r^2+a^2)+\tilde P]/\Delta$ with
$\tilde P=\sqrt{(r^2+a^2)^2-\Delta\tilde K}$, so both $\Delta$ and $\tilde P$
carry $M(u)$. Differentiating at fixed $r$ and $\tilde K$ gives
\begin{equation}
\partial_MA_V=\frac{r\tilde K}{\tilde P\Delta}
+\frac{2r\bigl[(r^2+a^2)+\tilde P\bigr]}{\Delta^{2}}
\;\longrightarrow\;\frac{4r(r^2+a^2)}{\Delta^{2}},
\label{eq:dAV}
\end{equation}
where the limit is the near-horizon limit. Since $A_V$ is the amount by which the
ingoing null coordinate advances per unit $r$, its mass dependence means that a
fixed radius sits at a different null-coordinate distance from the horizon as
the horizon shrinks, and the drift is the integral of that change,
\begin{equation}
\left.\frac{\partial V}{\partial u}\right|_{r}
=e^{\lambda_V}\Bigl[1+\int_{r_{\rm ref}}^{r}\partial_MA_V\,M'\,\dd r'\Bigr].
\label{eq:drift}
\end{equation}
The factor $\Delta^{-2}$ in \eqref{eq:dAV} is the reason why the drift is
largest where a curve leaves the horizon, which is the region magnified in
Fig.~\ref{fig:bend}. Because $V$ is normalised at $r=10$, the integral in
\eqref{eq:drift} changes sign there, so the curves with $r<10$ drift upwards
while those with $r>10$ drift downwards, as the figure shows. This drift should
not be confused with the behaviour of the Carter function in
\eqref{eq:kt-exp}, which originates in the $M'$ term of the ingoing master
equation rather than in $A_V$.

\section{Conclusion and discussion}\label{sec:conclusion}
We have constructed a pair of null coordinates for the Kerr-Vaidya spacetime. The construction requires two independent Carter-type functions $K(u,r,\theta)$ and $\tilde K(u,r,\theta)$, one for each null congruence, together with the hemisphere sign $\sigma(\theta)=\operatorname{sign}(\cos\theta)$ and two integrating factors $\lambda$ and $\lambda_V$.  The closure conditions for the outgoing and ingoing canonical one-forms, Eqs.~\eqref{eq:M1} and \eqref{eq:M2}, are different equations, and requiring them to hold with the same function would impose the additional constraint \eqref{eq:singleK}, which is incompatible with the master equation except in the static limit. For $M'=0$ both equations reduce to the Arga\~naraz--Moreschi equation and $\tilde K=K$.

We derived the two master equations and the two integrating-factor quadratures, solved both sectors numerically by a method-of-lines scheme, and obtained the Kerr--Vaidya metric in double-null form, Eq.~\eqref{eq:kv-dn}. We also showed that the unextended null coordinates are singular at the horizon.  For a time-dependent mass the surface gravity is itself a function of the retarded time, and we show that the rescaled coordinates, with the surface gravity \eqref{eq:kappa}, are finite at the horizon and that the extended metric is regular there to leading adiabatic order.

The immediate applications of these coordinates are the study of the evolution of massless scalar fields in the Kerr-Vaidya spacetime and the construction of the Penrose-type global conformal diagram, as in \cite{Arganaraz2022,Arganaraz2023}. The study can also be extended to asymptotically Einstein, that is, non-flat, or cosmological backgrounds, since Kerr-Vaidya-type metrics are already known in these settings \cite{Ramachandra2009,VaidyaPatel1989}. In the stationary Vaidya--Einstein--Kerr background a genuine Carter constant does exist \cite{Ramachandra2009}, and it would be of interest to determine how it is deformed in the time-dependent case. A further extension is to Kerr-Vaidya metrics with a variable angular-momentum-to-mass ratio $a(u)$ \cite{Dahal2025,BH17}, for which the master equations acquire source terms proportional to $a'(u)$. We leave these problems for future work.

\begin{acknowledgments}
X.G. is supported by Yancheng Institute of Technology (xjr2024030). X.P. is supported by the National Natural Science Foundation of China (NSFC) with Grant No. 12505082, and by the Doctoral Initiation Grant 24KE051 and Basic Research Grant 25kx010 from China West Normal University. X.G. also thanks Ping Li for helpful discussions.
\end{acknowledgments}

\appendix
\section{Derivation of the Eddington--Finkelstein form and its inverse}\label{AnnA}
For completeness we collect the algebraic details of deriving the form from the Boyer-Lindquist form of the Kerr metric.

Consider the Eddington-Finkelstein transformation with a constant mass
\begin{equation}
\dd t=\dd u+\frac{r^{2}+a^{2}}{\Delta}\dd r,\qquad
\dd\phi=\dd\bar\varphi+\frac{a}{\Delta}\dd r .
\end{equation}
\begin{widetext}
Substituting the three quadratic forms
\begin{align}
\dd t^{2}&=\dd u^{2}+2\frac{r^{2}+a^{2}}{\Delta}\dd u\,\dd r+\frac{(r^{2}+a^{2})^{2}}{\Delta^{2}}\dd r^{2},\label{eq:dt2}\\
\dd t\,\dd\phi
&=\dd u\,\dd\bar\varphi+\frac{a}{\Delta}\dd u\,\dd r
+\frac{r^{2}+a^{2}}{\Delta}\dd r\,\dd\bar\varphi
+\frac{a(r^{2}+a^{2})}{\Delta^{2}}\dd r^{2},\label{eq:dtdphi}\\
\dd\phi^{2}&=\dd\bar\varphi^{2}+2\frac{a}{\Delta}\dd r\,\dd\bar\varphi+\frac{a^{2}}{\Delta^{2}}\dd r^{2},\label{eq:dphi2}
\end{align}
into the metric \eqref{eq:BL} and collecting coefficients term by term gives the following.

{\it Coefficient of $\dd u^{2}$.} Using $\Delta-a^{2}\sin^{2}\theta=r^{2}-2Mr+a^{2}-a^{2}\sin^{2}\theta=\rho^{2}-2Mr$, we get
\begin{equation}
-\frac{\Delta-a^{2}\sin^{2}\theta}{\rho^{2}}=-\Bigl(1-\frac{2Mr}{\rho^{2}}\Bigr).
\end{equation}
{\it Coefficient of $\dd u\,\dd r$.}
\begin{equation}
-\frac{2(r^{2}+a^{2})(\Delta-a^{2}\sin^{2}\theta)}{\rho^{2}\Delta}
-\frac{4Ma^{2}r\sin^{2}\theta}{\rho^{2}\Delta}
=-\frac{2(r^{2}+a^{2})\Delta-2a^{2}\sin^{2}\theta(r^{2}+a^{2}-2Mr)}{\rho^{2}\Delta}.
\end{equation}
Using $r^{2}+a^{2}-2Mr=\Delta$ this equals
$-\dfrac{2\Delta(r^{2}+a^{2}-a^{2}\sin^{2}\theta)}{\rho^{2}\Delta}
=-\dfrac{2\Delta(r^{2}+a^{2}\cos^{2}\theta)}{\rho^{2}\Delta}=-2$, so that $g_{ur}=-1$.

{\it Coefficient of $\dd r^{2}$.} Collecting the four contributions, we have
\begin{equation}
\frac{1}{\rho^{2}\Delta^{2}}\Bigl[
\rho^{4}\Delta-(\Delta-a^{2}\sin^{2}\theta)(r^{2}+a^{2})^{2}
-4Ma^{2}r\sin^{2}\theta(r^{2}+a^{2})+a^{2}\Sigma\sin^{2}\theta\Bigr].
\end{equation}
Writing $\Sigma=(r^{2}+a^{2})^{2}-a^{2}\Delta\sin^{2}\theta$, the bracket reduces to
\begin{equation}
\Delta\bigl[\rho^{4}-(r^{2}+a^{2})^{2}-a^{4}\sin^{4}\theta+2a^{2}(r^{2}+a^{2})\sin^{2}\theta\bigr].
\end{equation}
Using $\rho^{4}-(r^{2}+a^{2})^{2}=-a^{2}\sin^{2}\theta\bigl[2r^{2}+a^{2}(1+\cos^{2}\theta)\bigr]$, the
square bracket vanishes identically, hence $g_{rr}=0$. This cancellation is a purely algebraic identity in
$M$, so it survives the replacement $M\to M(u)$.

{\it Coefficient of $\dd u\,\dd\bar\varphi$.}
\begin{equation}
-\frac{4Mar\sin^{2}\theta}{\rho^{2}}\;\Longrightarrow\;g_{u\bar\varphi}=-\frac{2Mar\sin^{2}\theta}{\rho^{2}} .
\end{equation}

{\it Coefficient of $\dd r\,\dd\bar\varphi$.}
\begin{equation}
-\frac{4Mar\sin^{2}\theta(r^{2}+a^{2})}{\rho^{2}\Delta}
+\frac{2a\Sigma\sin^{2}\theta}{\rho^{2}\Delta}
=\frac{2a\sin^{2}\theta}{\rho^{2}\Delta}\bigl[(r^{2}+a^{2})\Delta-a^{2}\Delta\sin^{2}\theta\bigr]
=2a\sin^{2}\theta,
\end{equation}
so $g_{r\bar\varphi}=a\sin^{2}\theta$. The $\dd\bar\varphi^{2}$ and $\dd\theta^{2}$ terms are unchanged.

The determinant of the block \eqref{eq:block} is computed by cofactor expansion along the second column
(where $B_{22}=0$):
\begin{equation}
\det B=-g_{uu}g_{r\bar\varphi}^{2}-g_{\bar\varphi\bar\varphi}-2g_{r\bar\varphi}g_{u\bar\varphi} .
\end{equation}
Substituting $g_{uu}=-(1-2Mr/\rho^{2})$, $g_{r\bar\varphi}=a\sin^{2}\theta$,
$g_{u\bar\varphi}=-2Mar\sin^{2}\theta/\rho^{2}$, $g_{\bar\varphi\bar\varphi}=\Sigma\sin^{2}\theta/\rho^{2}$, we have
\begin{equation}
\det B=\frac{a^{2}\sin^{4}\theta(\rho^{2}-2Mr)}{\rho^{2}}
-\frac{\Sigma\sin^{2}\theta}{\rho^{2}}
+\frac{4Ma^{2}r\sin^{4}\theta}{\rho^{2}}
=\frac{\sin^{2}\theta}{\rho^{2}}\Bigl[a^{2}\sin^{2}\theta(\rho^{2}+2Mr)-\Sigma\Bigr].
\end{equation}
Using the identity $\Sigma=\rho^{4}+a^{2}\sin^{2}\theta(\rho^{2}+2Mr)$, one finds
\begin{equation}\det B=-\rho^{2}\sin^{2}\theta.\end{equation}
Then
\begin{equation}g^{uu}=\mathrm{cofactor}_{uu}/\det B=-g_{r\bar\varphi}^{2}/\det B
=a^{2}\sin^{2}\theta/\rho^{2}.\end{equation}

\section{Ingoing Kerr--Vaidya case}\label{app:ing}
The ingoing metric ($M=M(v)$) is
\begin{equation}
\dd s^{2}=-\Bigl(1-\frac{2Mr}{\rho^{2}}\Bigr)\dd v^{2}+2\dd v\,\dd r
-\frac{4Mar\sin^{2}\theta}{\rho^{2}}\dd v\,\dd\bar\varphi
-2a\sin^{2}\theta\,\dd r\,\dd\bar\varphi+\rho^{2}\dd\theta^{2}
+\frac{\Sigma\sin^{2}\theta}{\rho^{2}}\dd\bar\varphi^{2},
\label{eq:ing-metric}
\end{equation}
with inverse
\begin{equation}
g^{vv}=\frac{a^{2}\sin^{2}\theta}{\rho^{2}},\quad
g^{vr}=\frac{r^{2}+a^{2}}{\rho^{2}},\quad
g^{v\bar\varphi}=\frac{a}{\rho^{2}},\quad
g^{rr}=\frac{\Delta}{\rho^{2}},\quad
g^{r\bar\varphi}=\frac{a}{\rho^{2}},\quad
g^{\theta\theta}=\frac{1}{\rho^{2}},\quad
g^{\bar\varphi\bar\varphi}=\frac{1}{\rho^{2}\sin^{2}\theta}.
\label{eq:ing-inv}
\end{equation}
The axisymmetric eikonal equation reads
\begin{equation}
a^{2}\sin^{2}\theta(\partial_{v}V)^{2}
+2(r^{2}+a^{2})\partial_{v}V\,\partial_{r}V
+\Delta(\partial_{r}V)^{2}+(\partial_{\theta}V)^{2}=0,
\label{eq:ing-eik}
\end{equation}
differing from \eqref{eq:eikonal} only in the sign of the $\partial_{v}V\,\partial_{r}V$ term. With
$\partial_{v}V=1$ and $(\partial_{\theta}V)^{2}=\tilde K-a^{2}\sin^{2}\theta$, the radial roots are
\begin{equation}
\partial_{r}V=\frac{-(r^{2}+a^{2})\pm\tilde P}{\Delta},
\end{equation}
i.e.\ the ingoing branch $\partial_{r}V=-A$ and the outgoing branch $\partial_{r}V=-A_V$ with
$A_V=\bigl[(r^{2}+a^{2})+\tilde P\bigr]/\Delta$. As $a\to0$ (with $\tilde K\to0$, $\tilde P\to r^{2}$), the ingoing branch
tends to $\dd v$ and the outgoing branch to $\dd v-\dfrac{2}{1-2M/r}\dd r$, the standard Vaidya null pair of
Sec.~\ref{sec:vaidya}.

It remains to justify the relative sign of the $\dd\theta$ components of the two one-forms. For a general one-form
$\varpi=\dd u+\mathcal A\dd r+\mathcal B\dd\theta$ the closure condition in the static case is
$\mathcal B_{r}=\mathcal A_{\theta}$. For the outgoing one-form, $\mathcal B=B=-\sigma h$ and
$\mathcal A=A$, so that $-\sigma\partial_{r}K/(2h)=\partial_{\theta}K/(2P)$, i.e.\
$P\partial_{r}K+\sigma h\partial_{\theta}K=0$. For the ingoing one-form with the \emph{same} $\tilde K=K$ and with
$\mathcal B=B_V=+\sigma h$, $\mathcal A=A_V$, the condition $B_{V,r}=A_{V,\theta}$ gives
$\sigma\partial_{r}K/(2h)=-\partial_{\theta}K/(2P)$, i.e.\ again $P\partial_{r}K+\sigma h\partial_{\theta}K=0$.
Had we chosen the same sign for the two $\dd\theta$ components, the two conditions would have differed by the sign of
$\partial_{\theta}K$ and could not both be satisfied. This is why $B$ and $B_V$ carry opposite signs, and why the
two congruences are described by the same function in the static case.

The analysis of Sec.~\ref{sec:master} applies, with the obvious modifications, to the
ingoing metric \eqref{eq:ing-metric}: its null one-forms are obtained from
\eqref{eq:omega-omegaV} by $u\to v$ and by the sign changes recorded above, and their
closure conditions lead to master equations of the same structure as \eqref{eq:M1}
and \eqref{eq:M2}. We do not repeat the construction here.

\section{Derivatives of $P$, $A$, $B$}\label{app:deriv}
From $P^{2}=(r^{2}+a^{2})^{2}-\Delta K$, we can compute
\begin{equation}
\partial_{K}P=-\frac{\Delta}{2P},\qquad
\partial_{\Delta}P=-\frac{K}{2P},\qquad
\partial_{\theta}P=-\frac{\Delta\,\partial_{\theta}K}{2P},\qquad
\partial_{u}P=\frac{2rM'(u)K-\Delta\,\partial_{u}K}{2P},
\label{eq:dP}
\end{equation}
using $\partial_{u}\Delta=-2rM'(u)$. With $A=\bigl[(r^{2}+a^{2})-P\bigr]/\Delta=K/\bigl[P+(r^{2}+a^{2})\bigr]$, we have
\begin{equation}
\partial_{K}A=\frac{1}{2P},\qquad
\partial_{\Delta}A=\frac{K}{2P\Delta}-\frac{(r^{2}+a^{2})-P}{\Delta^{2}}
=\frac{\bigl[(r^{2}+a^{2})-P\bigr]^{2}}{2P\Delta^{2}}=\frac{A^{2}}{2P}.
\label{eq:dA}
\end{equation}
Hence
\begin{equation}
\partial_{u}A=\partial_{K}A\,\partial_{u}K+\partial_{\Delta}A\,\partial_{u}\Delta
=\frac{\partial_{u}K}{2P}-\frac{2rM'(u)A^{2}}{2P}
=\frac{\partial_{u}K-2rM'(u)A^{2}}{2P},
\end{equation}
and similarly
\begin{equation}
\partial_{\theta}A=\frac{\partial_{\theta}K}{2P},\qquad
\partial_{r}B=-\sigma\frac{\partial_{r}K}{2h},\qquad
\partial_{u}B=-\sigma\frac{\partial_{u}K}{2h},
\end{equation}
with $h=\sqrt{K-a^{2}\sin^{2}\theta}$. The identity $\partial_{\Delta}A=A^{2}/(2P)$ is proven as follows:
\begin{equation}
\frac{K}{2P\Delta}-\frac{(r^{2}+a^{2})-P}{\Delta^{2}}
=\frac{K\Delta-2P(r^{2}+a^{2})+2P^{2}}{2P\Delta^{2}}
=\frac{(r^{2}+a^{2})^{2}-P^{2}-2P(r^{2}+a^{2})+2P^{2}}{2P\Delta^{2}}
=\frac{\bigl[(r^{2}+a^{2})-P\bigr]^{2}}{2P\Delta^{2}},
\end{equation}
using $K\Delta=(r^{2}+a^{2})^{2}-P^{2}$.

For the ingoing coefficients, the same manipulations with $\tilde P$ and $\tilde K$ give
\begin{equation}
\partial_{\tilde K}\tilde P=-\frac{\Delta}{2\tilde P},\qquad
\partial_{\Delta}\tilde P=-\frac{\tilde K}{2\tilde P},\qquad
\partial_{\theta}\tilde P=-\frac{\Delta\,\partial_{\theta}\tilde K}{2\tilde P},\qquad
\partial_{u}\tilde P=\frac{2rM'\tilde K-\Delta\,\partial_{u}\tilde K}{2\tilde P},
\end{equation}
and, with $A_V=\bigl[(r^{2}+a^{2})+\tilde P\bigr]/\Delta$,
\begin{equation}
\partial_{\tilde K}A_V=-\frac{1}{2\tilde P},\qquad
\partial_{\Delta}A_V=-\frac{A_V^{2}}{2\tilde P},\qquad
\partial_{\theta}A_V=-\frac{\partial_{\theta}\tilde K}{2\tilde P},\qquad
\partial_{r}B_V=+\sigma\frac{\partial_{r}\tilde K}{2\tilde h},\qquad
\partial_{u}B_V=+\sigma\frac{\partial_{u}\tilde K}{2\tilde h},
\end{equation}
so that
\begin{equation}
\partial_{u}A_V=-\frac{\partial_{u}\tilde K}{2\tilde P}+\frac{rM'(u)A_V^{2}}{\tilde P},
\qquad \tilde h=\sqrt{\tilde K-a^{2}\sin^{2}\theta}.
\end{equation}
Note the opposite signs with respect to \eqref{eq:dA}. They are responsible for the different
coefficients of $\partial_u\tilde K$ and $M'$ in \eqref{eq:M2}.

\section{Why a single Carter-type function is insufficient}\label{app:singleK}
In this Appendix we derive the constraint \eqref{eq:singleK}. Set $\tilde K=K$, so that $\tilde P=P$,
$\tilde h=h$ and $A_V=(\rho_1+P)/\Delta$, and subtract \eqref{eq:M1} from \eqref{eq:M2}. The left-hand sides
coincide, while the right-hand sides differ by
\begin{equation}
\Bigl(\frac{A_V\partial_uK}{h}+\frac{h\partial_uK}{P}-\frac{2rhM'A_V^{2}}{P}\Bigr)
-\Bigl(\frac{A\partial_uK}{h}-\frac{h\partial_uK}{P}+\frac{2rhM'A^{2}}{P}\Bigr)
=\frac{(A_V-A)\partial_uK}{h}+\frac{2h\partial_uK}{P}-\frac{2rhM'(A_V^{2}+A^{2})}{P}.
\end{equation}
Using $A_V-A=2P/\Delta$ and $A_V^{2}+A^{2}=2(\rho_1^{2}+P^{2})/\Delta^{2}$, the condition that both
equations hold becomes
\begin{equation}
\frac{2P\partial_uK}{\Delta h}+\frac{2h\partial_uK}{P}=\frac{4rhM'(\rho_1^{2}+P^{2})}{\Delta^{2}P},
\end{equation}
i.e.
\begin{equation}
\partial_uK\,\frac{P^{2}+\Delta h^{2}}{P\Delta h}=\frac{2rhM'(\rho_1^{2}+P^{2})}{\Delta^{2}P}.
\end{equation}
Since $P^{2}+\Delta h^{2}=(r^{2}+a^{2})^{2}-\Delta K+\Delta K-\Delta a^{2}\sin^{2}\theta=\Sigma$, this is
\begin{equation}
\partial_uK=\frac{2M'(u)\,r\,h^{2}\bigl(\rho_1^{2}+P^{2}\bigr)}{\Sigma\Delta},
\end{equation}
which is Eq.~\eqref{eq:singleK}. For $M'=0$ it reduces to $\partial_uK=0$, which is compatible with the
static master equation \eqref{eq:AM}. For $M'\neq0$ it is an additional algebraic constraint on $K$, since
\eqref{eq:M1} already determines $\partial_uK$, and a generic solution of the outgoing master equation does not
satisfy it. Hence a single function cannot in general close both one-forms, and the two functions
$K$ and $\tilde K$ must be kept independent.

\section{The double-null form of the metric}\label{AnnD}
Here we verify the double-null form \eqref{eq:kv-dn} and the statements made in Sec.~\ref{sec:dn}.

{\it Derivation.} Equations \eqref{eq:app-dr} and \eqref{eq:app-coeffs} are obtained by solving
\eqref{eq:omega-omegaV} for $\dd r$ and $\dd u$:
\begin{equation}
\dd r=\frac{\omega-\omega_{V}-(B-B_V)\dd\theta}{A-A_V}
=\frac{\Delta}{P+\tilde P}\bigl(\omega_{V}-\omega\bigr)+\frac{\Delta(B-B_V)}{P+\tilde P}\dd\theta,
\end{equation}
since $A-A_V=-(P+\tilde P)/\Delta$, and then $\dd u=\omega-A\dd r-B\dd\theta$. Substituting these
expressions into \eqref{eq:kv-metric}, the coefficients of the ten independent quadratic terms are
\eqref{eq:kv-dn-coeffs}. For instance the coefficient of $\omega^{2}$ receives the contributions
$g_{uu}\alpha^{2}+2g_{ur}\alpha(-\delta)=-f\alpha^{2}+2\alpha\delta$, and the coefficient of
$\dd\theta\,\dd\bar\varphi$ is $2g_{u\bar\varphi}\gamma+2g_{r\bar\varphi}\varepsilon
=-2w\gamma+2j\varepsilon=2\mathfrak I$, which is the origin of the cross term that cannot be removed
when $M'\neq0$.

{\it The inverse metric and $g^{UU}=g^{VV}=0$.} It is more compact to transform the inverse metric. With
\begin{equation}
\frac{\partial U}{\partial u}=e^{\lambda},\quad
\frac{\partial U}{\partial r}=e^{\lambda}A,\quad
\frac{\partial U}{\partial\theta}=e^{\lambda}B,\qquad
\frac{\partial V}{\partial u}=e^{\lambda_V},\quad
\frac{\partial V}{\partial r}=e^{\lambda_V}A_V,\quad
\frac{\partial V}{\partial\theta}=e^{\lambda_V}B_V,
\end{equation}
and $g^{\theta\theta}=1/\rho^{2}$, $g^{\bar\varphi\bar\varphi}=1/(\rho^{2}\sin^{2}\theta)$,
$g^{u\bar\varphi}=a/\rho^{2}$, $g^{r\bar\varphi}=-a/\rho^{2}$, one obtains
\begin{align}
&g^{UU}=g^{VV}=0,\qquad
g^{UV}=\frac{e^{\lambda+\lambda_V}}{\rho^{2}}\,\mathcal N,\qquad
g^{U\theta}=\frac{e^{\lambda}B}{\rho^{2}},\qquad
g^{V\theta}=\frac{e^{\lambda_V}B_V}{\rho^{2}},\nonumber\\
&g^{U\bar\varphi}=\frac{e^{\lambda}a(1-A)}{\rho^{2}},\qquad
g^{V\bar\varphi}=\frac{e^{\lambda_V}a(1-A_V)}{\rho^{2}},\qquad
g^{\theta\theta}=\frac{1}{\rho^{2}},\qquad
g^{\theta\bar\varphi}=0,\qquad
g^{\bar\varphi\bar\varphi}=\frac{1}{\rho^{2}\sin^{2}\theta},
\label{eq:inv-dn}
\end{align}
with
\begin{equation}
\mathcal N\equiv a^{2}\sin^{2}\theta-\rho_1\bigl(A+A_V\bigr)+\Delta AA_V+BB_V .
\end{equation}
The vanishing of $g^{UU}$ is precisely the null condition \eqref{eq:nullcheck}. In the static limit
$\mathcal N$ reduces to $-2\Sigma/\Delta$, which is nonvanishing for $\Delta>0$, so that $g^{UV}\neq0$
and the two null directions are independent. The divergence of $\mathcal N$ as $\Delta\to0$ is the
degeneration of the chart discussed in Sec.~\ref{app:dn-horizon}.

{\it Determinant.} Since the transformation $(U,V)\to(u,r)$ at fixed $(\theta,\bar\varphi)$ has Jacobian
$J=e^{\lambda+\lambda_V}(P+\tilde P)/\Delta$, one has
$\det g=\det g_{\rm KV}/J^{2}$, and with
$\det g_{\rm KV}=-\rho^{4}\sin^{2}\theta$ (cf.\ Appendix~\ref{AnnA}) this gives
\begin{equation}
\det g=-\frac{\rho^{4}\sin^{2}\theta\,\Delta^{2}}{(P+\tilde P)^{2}}e^{-2(\lambda+\lambda_V)},
\end{equation}
which is Eq.~\eqref{eq:det-dn}. Equivalently, this follows from \eqref{eq:inv-dn} by direct computation.

\end{widetext}

\end{document}